\documentclass[aps,pre,bibnotes,superscriptaddress]{revtex4}

\usepackage{graphicx}
\usepackage{amssymb}
\usepackage{amsmath}
\usepackage{amsfonts}
\usepackage{hyperref}
\usepackage{bm}

\DeclareMathAccent{\ring}{\mathalpha}{operators}{"17}
\providecommand{\st}[1]{_{\text{#1}}}

\providecommand{\ten}[1]{{\bv{#1}}}

\def\onehalf{\frac{1}{2}}

\def\bra{\ensuremath{\langle}}
\def\ket{\ensuremath{\rangle}}
\def\eq{\st{eq}}

\def\pd{\partial}

\def\im{\mathrm{i}}

\def\kv{\bv{k}}

\def\uv{\bv{u}}

\def\rv{\bv{r}}

\def\b0{\bv{0}}

\def\ra{\rightarrow}

\def\Fcal{\mathcal{F}}

\def\wdmp{\text{wd}}
\def\sdmp{\text{sd}}

\newcommand{\bitem}{\begin{itemize}}
\newcommand{\eitem}{\end{itemize}}
\newcommand{\benum}{\begin{enumerate}}
\newcommand{\eenum}{\end{enumerate}}
\newcommand{\bblock}[1]{\begin{block}{#1}}
\newcommand{\eblock}{\end{block}}
\newcommand{\bmini}[1]{\begin{minipage}{#1}}
\newcommand{\emini}{\end{minipage}}
\newcommand{\btab}[1]{\begin{tabular}{#1}}
\newcommand{\etab}{\end{tabular}}
\newcommand{\btabn}[1]{\begin{tabular}{#1}}
\newcommand{\etabn}{\end{tabular}}
\newcommand{\beq}{\begin{equation}}
\newcommand{\eeq}{\end{equation}}
\newcommand{\beqn}{\begin{equation*}}
\newcommand{\eeqn}{\end{equation*}}
\newcommand{\bmult}{\begin{multline}}
\newcommand{\emult}{\end{multline}}
\newcommand{\bsplit}{\begin{split}}
\newcommand{\esplit}{\end{split}}

\newcommand{\bv}[1]{\mathbf{#1}}

\begin{document}
 \title{Interfacial roughening in non-ideal fluids:\\Dynamic scaling in the weak- and strong-damping regime}
 \author{Markus Gross}
 \email{markus.gross@rub.de}
 \affiliation{Interdisciplinary Centre for Advanced Materials Simulation (ICAMS), Ruhr-Universit\"at Bochum, Universit\"atsstr. 90a, 44789 Bochum, Germany}
 \author{Fathollah Varnik}
 \affiliation{Interdisciplinary Centre for Advanced Materials Simulation (ICAMS), Ruhr-Universit\"at Bochum, Universit\"atsstr. 90a, 44789 Bochum, Germany}
 \affiliation{Max-Planck Institut f\"ur Eisenforschung, Max-Planck Str.~1, 40237 D\"usseldorf, Germany}

\begin{abstract}
Interfacial roughening denotes the nonequilibrium process by which an initially flat interface reaches its equilibrium state, characterized by the presence of thermally excited capillary waves. 
Roughening of fluid interfaces has been first analyzed by Flekkoy and Rothman [Phys. Rev. Lett. {\bf 75}, 260 (1995)], where the dynamic scaling exponents in the weakly damped case in two dimensions were found to agree with the Kardar-Parisi-Zhang universality class. 
We extend this work by taking into account also the strong-damping regime and perform extensive fluctuating hydrodynamics simulations in two dimensions using the Lattice Boltzmann method. We show that the dynamic scaling behavior is different in the weakly and strongly damped case. 
\end{abstract}


\maketitle

\section{Introduction}
Capillary fluctuations on an interface between two fluid phases are waves that are excited by thermal noise in the bulk \cite{mandelstam_1913, buff_1965, evans_interface_1979,RowlinsonWidom_book, safran_book, loudon_1980, grant_desai_1983}.
From a macroscopic viewpoint, capillary fluctuations make the interface ``rough'' and increase the effective interface width.
Roughening of interfaces is a fundamental aspect of nonequilibrium dynamics and has been widely studied in the literature (see, e.g., \cite{family_viscek_book1991, barabasi_stanley_fractal_book, krug_spohn_roughening_1991, meakin_physrep1993, halpinhealy_zhang_roughening_1995} for reviews). 
Most models for interface growth, such as the Edwards-Wilkinson \cite{edwards_wilkinson_1982} or Kardar-Parisi-Zhang \cite{kpz_prl1986} equations, describe a purely local growth mechanism. In the case of an interface between two fluids, however, one can expect significant dynamical effects arising from the coupling of the order parameter to the hydrodynamic flow field. Indeed, as has been shown in \cite{flekkoy_1995, flekkoy_1996}, the effective Langevin description of a roughening fluid interface is in general non-Markovian due to the surrounding flow.
The roughening of fluid interfaces due to thermal fluctuations is potentially relevant for the stability of patterns that form, for instance, in reaction-diffusion systems \cite{foard_wagner_morphologies_pre2012, ayodele_grayscott_pre2011} or during phase-transitions under shear \cite{wagner_yeomans_shear_pre1999, shou_chakrabarti_pre2000, bray_short_pre2001, bray_long_pre2001}. Also, coalescence of droplets or films \cite{aarts_coalescence_jfm2008} or the dynamics of wetting transitions \cite{grant_wetting_prb1988} are potentially affected by interfacial roughening.

Thermal roughening of fluid interfaces was first studied in \cite{flekkoy_1995, flekkoy_1996} based on the equations of fluctuating hydrodynamics and by simulations of an immiscible lattice gas (see also \cite{starr_boghosian_prl1996} for the same problem in the presence of surfactants). There, it was concluded that the roughening dynamics of a weakly damped interface is characterized by the scaling exponents of the Kardar-Parisi-Zhang \cite{kpz_prl1986} universality class.
In the present work, interfacial roughening of small-amplitude capillary waves is investigated in the strong-damping regime and the associated dynamic scaling exponents and scaling forms are derived. We show that the growth of the interfacial roughness in the strong-damping regime is qualitatively and quantitatively different from the weakly damped case. The theoretical predictions are compared against fluctuating hydrodynamics simulations of an isothermal liquid-vapor interface.

The paper is organized as follows: In the next section, the Langevin approach to capillary fluctuations of \cite{flekkoy_1995, flekkoy_1996} is summarized and applied to the roughening dynamics in the strong-damping regime. 
Section \ref{sec:sim} contains results of fluctuating hydrodynamics simulations of a single-component two-phase fluid performed with the Lattice Boltzmann method. After demonstrating that both static and dynamic equilibrium properties of capillary waves are correctly reproduced by the simulations, the non-equilibrium roughening of a fluid interface is investigated and compared to the theoretical predictions. 

\section{Theory}
\subsection{General description of capillary waves}

\begin{figure}[t]\centering
    \includegraphics[width=0.37\linewidth]{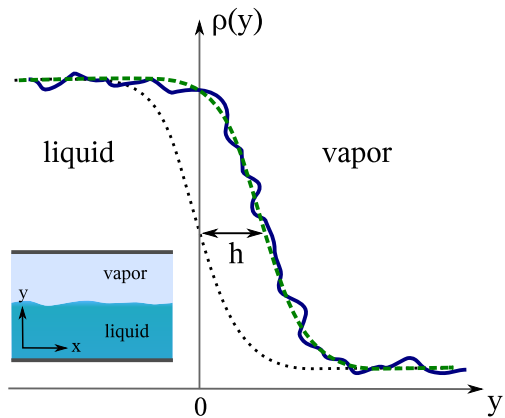}\quad
   \caption{(Color online) Capillary fluctuations of a planar interface. Sketch of the fluctuating density profile as a function of the lateral coordinate $y$. The dotted curve represents the quiescent, mean-field profile, while the dashed curve represents the fit of the mean-field profile to the instantaneous density profile (solid curve). The inset shows the simulation setup and coordinate axis.}
    \label{fig:capill-def}
\end{figure}

Capillary waves can be described in terms of a local height function $h(\rv_{||})$, where $\rv_{||}$ denotes a position in the interfacial plane (Fig.~\ref{fig:capill-def}). The projected area of the interface is given by $L^{d-1}$. In the two-dimensional situation we focus upon, $r_{||} = x$, while the perpendicular coordinate is denoted by $y$.
In the classical capillary wave theory \cite{mandelstam_1913, buff_1965, evans_interface_1979, RowlinsonWidom_book}, a capillary fluctuation is understood as a rigid shift of the ``intrinsic'' density profile. Thus, we can define the height function $h$ as
\beq \rho(\rv) = \rho\st{int}(y-h(\rv_{||}))\,,\label{int_height_def} \eeq
where $\rho\st{int}$ is the intrinsic and $\rho(\rv)$ the instantaneous density profile.
Since, in the present case, our treatment of a two-phase fluid is based on a Ginzburg-Landau model (see sec.~\ref{sec:model}), we can take as the intrinsic profile the mean-field solution
\beq \rho\st{int}(y) = \onehalf (\rho_L+\rho_V) + \onehalf (\rho_L-\rho_V) \tanh\left(\frac{y}{w}\right)\,,
\label{intprof}
\eeq
where $\rho_L$ and $\rho_V$ are the liquid and vapor densities, respectively, and $w$ is the (bare) interface width.

In our simulations, we obtain the interfacial height $h$ by fitting $\rho\st{int}$ to the instantaneous density profile (see Fig.~\ref{fig:capill-def}).
If the interfacial height would, instead, be determined by means of a simple crossing criterion [i.e., $\rho(h)=(\rho_L+\rho_V)/2$], one would pick up local density fluctuations that are present in the interface due to its finite width. These should, however, not be interpreted as capillary waves since they are not associated with a lateral displacement of the interface profile, as expressed through eq.~\eqref{int_height_def}. Local density fluctuations affect the small-scale properties of the interfacial structure and can be understood in terms of interfacial density correlation functions \cite{zittartz_1967, diehl_zphysb1980, evans_1981, stecki_contrib_1998, sedlmeier_netz_prl2009, blokhuis_jcp2009}. In this work, however, we shall not consider them further \footnote{See, for instance, Fig. 5 in ref.~\cite{gross_flbe_2010} for an example of a capillary wave spectrum that results when the height profile is obtained from a crossing criterion} and stick to the definition of eq.~\eqref{int_height_def}. 

\subsection{Langevin theory}
For the description of non-equilibrium roughening, a Langevin approach to the capillary fluctuation dynamics is convenient. Such a formalism can be derived from the equations of fluctuating hydrodynamics \cite{grant_desai_1983, flekkoy_1995, flekkoy_1996}, and we shall base our treatment on the Langevin theory of \cite{flekkoy_1995, flekkoy_1996}, which is summarized below.
The general form of the Langevin equation for the height fluctuations $h_\kv$ of an infinitely deep film turns out to be non-Markovian \cite{flekkoy_1995, flekkoy_1996},
\beq h_\kv(t) = \int_{-\infty}^t ds \chi_\kv(t-s) F_\kv(s)\,,\label{cap_langevin}\eeq
where $\chi_\kv$ is a response function that, in Fourier space, is given by \cite{flekkoy_1995, flekkoy_1996}
\beq \chi_\kv(\omega) = \frac{1}{-\im \omega \gamma_\kv(\omega) + \sigma k^2},\quad \text{with}\quad \gamma_\kv(\omega) = \frac{2\omega \rho}{k\left(\im + \left(\frac{\im\omega}{\nu k^2}-1\right)^{-1/2}\right)}\, \label{cap_response}\eeq
and $F_\kv$ is a random force that satisfies a fluctuation-dissipation relation
\beq \bra F_\kv(t) F_\kv^*(0)\ket = k_B T \gamma_\kv(|t|)\,.\label{cap_random_force}\eeq
Physically, the random force $F_\kv$ arises from the accumulative effect of the random stress fluctuations in the bulk fluid up to a certain depth below the interface (cf.~\cite{grant_desai_1983}).
In the above equations, $\sigma$ denotes the surface tension and $\nu$ the kinematic viscosity of the liquid. In this work, we adopt the Fourier-transform convention $a(\rv,t) = (2\pi)^{-d} \int d\kv d\omega \exp(\im\kv\cdot \rv - \im \omega t) a_\kv(\omega)$, where $\kv$ denotes the wavevector in the $d-1$-dimensional interfacial plane ($d$ is the spatial dimension) and $\omega$ is a frequency.

In the limit of weak damping ($\nu k^2\ra 0$) one obtains, by expanding the square-root in $\gamma_\kv$ of eq.~\eqref{cap_response},
\beq \gamma_{\kv,\text{wd}}(\omega) = -\frac{2 \rho \im \omega }{k} - 2\rho \im\sqrt{\im\omega\nu} + 2\rho \nu k + O(\nu^{3/2}) \,.\label{cap_gamma_wd}\eeq
Keeping in the expansion of $\gamma_{\kv,\wdmp}$ only the first and the third term, the response function in the weak-damping limit acquires a harmonic oscillator form and is obtained from eq.~\eqref{cap_response} as
\beq \chi_{\kv,\wdmp}(\omega) = \frac{k}{2\rho} \frac{1}{-\omega^2 -\im \omega \nu k^2 + \frac{\sigma k^3}{2\rho}}\,.
\label{cap_resp_wd}
\eeq
In the time domain, this results in the Langevin equation
\beq -\pd_t^2 h_\kv + \nu k^2 \pd_t h_\kv + \frac{\sigma k^3}{2\rho} h_\kv = r_\kv\,,
\label{langevin_wd}
\eeq
with $r_\kv$ being a Gaussian random noise source with variance $\bra r_\kv(t) r_\kv^*(0)\ket =  k_B T [k \delta'(t) + \nu k^3 \delta(t)] /\rho$. 
The contribution $\propto \delta'(t)$ to the random force in the time-domain arises from the first term in eq.~\eqref{cap_gamma_wd} and is, therefore, present even in the absence of viscosity. This is one of the most characteristic features of the present Langevin theory, which, despite the principal harmonic oscillator form of the response function, leads to quite distinct noise-driven dynamics.
Equation~\eqref{langevin_wd} predicts capillary waves with an oscillation frequency and a damping rate given by
\beq \omega_c = \left(\frac{\sigma  k^3}{2\rho}\right)^{1/2}\,,\quad
\Gamma\st{wd} = \nu k^2 \,. \label{cap_disp}\eeq
The slight reduction of the resonance frequency $\omega_c$ due to a finite damping has been neglected here. 
Note that, if the second term in eq.~\eqref{cap_gamma_wd} also is kept, the damping rate would scale $\propto k^{7/4}$ \cite{flekkoy_1996}. Interestingly, this type of scaling has also been derived in a few previous works based on a different theoretical approach \cite{loudon_1980, jeng_capvisc_1998}. In our simulations, however, we observe a behavior in agreement with eq.~\eqref{cap_disp}, which can be rationalized in the context of the present Langevin theory only if the second term in eq.~\eqref{cap_gamma_wd} is neglected -- as we have done above. We also remark that the relations in eq.~\eqref{cap_disp} agree well with experiments \cite{madsen_prl2004} and other theoretical works \cite{harden_pleiner_pincus_jcp1991, jaeckle_kawasaki_jpcm1995}. While this fact provides some sort of justification of eq.~\eqref{cap_resp_wd}, further studies would be desirable in order to clarify the relevance of the additional terms in expression \eqref{cap_gamma_wd}. This, however, is out of the scope of the present work.

In the strong damping limit, one finds for $\nu k^2\ra \infty$:
\beq \gamma_{\kv,\text{sd}}(\omega) = 4\rho\nu k -\frac{3\rho\im \omega}{k} + \frac{\rho \omega^2}{4k^3\nu}+O(\nu^{-2})  \,.\label{cap_gamma_od}\eeq
Keeping only the leading term on the r.h.s.\ of eq.~\eqref{cap_gamma_od}, the response function follows as
\beq \chi_{\kv,\sdmp}(\omega) = \frac{1}{- 4\im \omega\rho\nu k  + \sigma k^2}\,,
\label{cap_resp_od}
\eeq
which results in the Langevin equation 
\beq \pd_t h_\kv + \frac{\sigma k}{4\rho\nu} h_\kv = \tilde r_\kv\,.
\label{langevin_sd}\eeq
Here, $\tilde r_\kv$ is a Gaussian random noise source with variance $\bra \tilde r_\kv(t) \tilde r_\kv^*(0)\ket = k_B T \delta(t) / 4\rho\nu k$.
Equation \eqref{langevin_sd} implies a decay rate of 
\beq \Gamma\st{sd} = \frac{\sigma k}{4 \rho \nu}\,.\label{cap_overdmp}\eeq
A Langevin equation for the height fluctuations of the form of eq.~\eqref{langevin_sd} has also been derived for $d>2$ in \cite{bray_long_pre2001, thiebaud_shear_pre2010}.

Weak and strong damping regimes are separated by a critical wavenumber $k_c$. An approximate value of $k_c$ can be obtained by noting that for small but finite damping, the capillary wave resonance in eq.~\eqref{cap_resp_wd} appears at a frequency of $(\omega_c^2-\Gamma\st{wd}^2/4)^{1/2}$, which becomes purely imaginary if $k>k_c$, where
\beq k_c=\frac{2\sigma}{\rho \nu^2}\,.
\eeq
Thus, the weak-damping regime applies to $k<k_c$ and the strong-damping regime to $k>k_c$. 
Clearly, the Langevin equation \eqref{langevin_sd} -- although with slightly different numerical prefactors -- could also have been directly obtained by taking the strong-damping limit of eq.~\eqref{langevin_wd}.

The dynamic correlation function $C(\kv,t)\equiv \bra h_{\kv}(t) h_{-\kv}(0)\ket$ follows directly from eq.~\eqref{cap_langevin}. The static correlation function can be most easily obtained from a fluctuation-response relation as
\beq C(\kv)\equiv \bra |h_{\kv}|^2\ket = k_B T \chi(\kv,\omega=0) = \frac{k_B T}{\sigma k^2}\,.
\label{cap_hcorr_eq}
\eeq
This classical result of capillary wave theory can also be derived from purely geometric considerations of the energy cost associated with a small-amplitude interfacial distortion \cite{RowlinsonWidom_book,safran_book}. In fact, the above result is equivalent to a harmonic approximation to the interface Hamiltonian, thus describing independent capillary waves. This is a valid approximation in the limit of small-amplitudes and large wavelengths.
In the presence of gravity, a finite correlation length is introduced into the static structure factor, thereby cutting off the divergence at low $k$ \cite{evans_interface_1979,RowlinsonWidom_book,safran_book}. 
It is useful to remark that the effective capillary wave Hamiltonian can also be obtained from a spectral analysis of a Ginzburg-Landau type of free energy functional \cite{zittartz_1967, evans_1981}. Such an approach has the advantage that, in principle, the density-correlation function in the interface can be derived from first-principles.

Up to numerical prefactors of the order of unity, the above expressions for the oscillation frequency and damping agree with the results of most theories of capillary wave dynamics in the literature \cite{levich_book_1962, bouchiat_meunier_jphys1972, grant_desai_1983, harden_pleiner_pincus_jcp1991, jaeckle_kawasaki_jpcm1995, mora_daillant_epjb2002}. In these theories, the liquid-vapor interface is usually taken as infinitesimally thin.
The expression for $\omega_c$ in eq.~\eqref{cap_disp} has also been explicitly derived for an inviscid fluid coupled to a Ginzburg-Landau free energy functional \cite{felderhoff_physica1970}.
Also, generalizations of capillary wave theory to non-zero vapor density \cite{bouchiat_meunier_jphys1974, jeng_capvisc_1998} as well as to compressible fluids \cite{loudon_1980, mora_daillant_epjb2002, falk_mecke_jpcm2011} have been proposed in the literature.
In the case of a compressible fluid, one finds that sound waves propagating parallel to the interface give rise to an additional resonance peak in the dynamic capillary structure factor at a frequency $\omega_s \simeq c_s k$. In the present study, this frequency is typically much larger than the resonance frequency of a capillary wave and will thus be neglected.

\subsection{Interfacial roughening}
\label{sec:roughening}

The effective \textit{interfacial roughness} is defined as the mean-square of the height amplitudes:
\beq W^2(t) = \bra |h(\rv,t)|^2\ket  = \frac{1}{L^{d-1}}\sum_\kv \bra |h_\kv(t)|^2\ket = \int \frac{d\kv}{(2\pi)^{d-1}} \bra |h_\kv(t)|^2\ket \,,
\label{cap_rough_def}
\eeq
which in equilibrium in given by \cite{mandelstam_1913, buff_1965, evans_interface_1979,RowlinsonWidom_book, flekkoy_1996},
\beq W^2\st{eq} \equiv W^2(t\ra \infty)  \simeq \frac{k_B T}{\sigma} \times
  \begin{cases}
   \frac{1}{12}L\,, & \text{in 2D} \\
   \log(L/l_0)\,, & \text{in 3D}
  \end{cases}
\label{cap_rough_eq}
\eeq
where $l_0$ is the minimal length scale available in the system. In the 2D-case, the sum has been computed using the relation $\sum_{n=1}^\infty 1/n^2 = \pi^2/6$, whereas in 3D, it has been approximated by an integral.
It is important to realize that, both in two and three dimensions, the interfacial roughness is diverging with the system size, although this divergence is very weak in 3D.
The time-evolution of the interface height under the action of the random force can be computed directly from the Langevin eq.~\eqref{cap_langevin}, explicitly assuming that $F_\kv(s)=0$ for $s<0$, since we are interested in the growth from a quiescent state at $t=0$. Note also that, due to causality, $\chi_\kv(t)=0$ for $t<0$.

\begin{figure}[t]\centering
    (a)\includegraphics[width=0.43\linewidth]{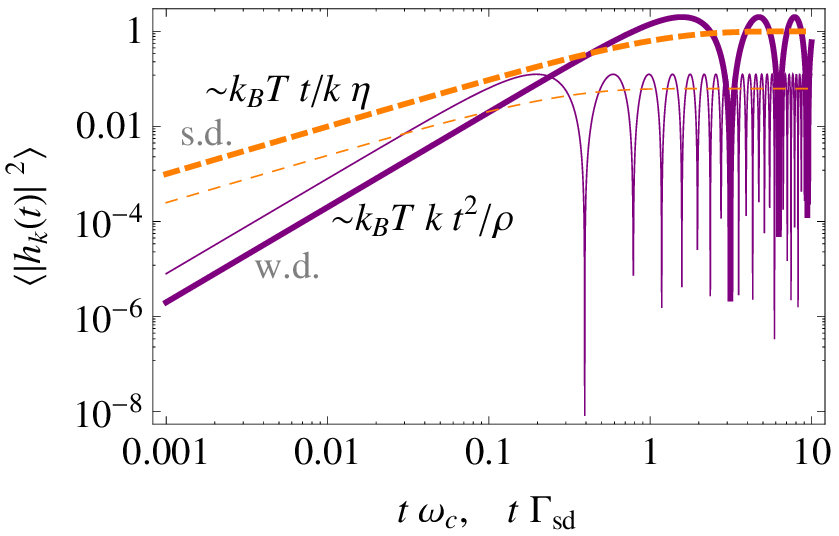}
    (b)\includegraphics[width=0.43\linewidth]{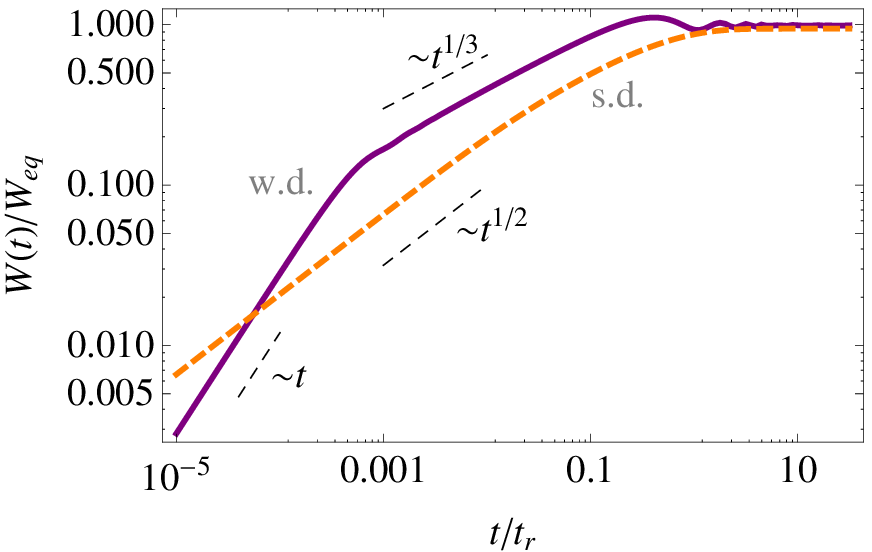}
    \caption{(Color online) Theoretical predictions of the growth of (a) the height amplitude $\bra |h_\kv(t)|^2\ket$ [eqs.~\eqref{cap_hcorr_wd} and \eqref{cap_hcorr_od}] and (b) the interfacial roughness $W(t)$ [eqs.~\eqref{cap_rough_wd} and \eqref{cap_rough_od}] in 2D. The solid and dashed curves correspond to the weak damping (`w.d.') and strong damping (`s.d.') regime, respectively. In (a), the thin curves additionally show the growth of a large wavenumber mode ($k'>k$), whose time and length is scaled with the same factors as the small wavenumber mode (thick curves) for better comparison. Time is scaled by the resonance frequency $\omega_c$ [eq.~\eqref{cap_disp}] or relaxation rate $\Gamma\st{sd}$ [eq.~\eqref{cap_overdmp}], depending on the regime. In (b), time is scaled by the corresponding roughening times, eqs.~\eqref{cap_tr_wd} and \eqref{cap_tr_od}.}
    \label{fig:cap_rough_th}
\end{figure}

Roughening in the \textit{weak-damping limit} has already been discussed in \cite{flekkoy_1995, flekkoy_1996}, of which the essential results shall be summarized first.
Neglecting the effect of viscosity, which is sufficient to determine the leading order behavior, the response function in the time-domain follows from eq.~\eqref{cap_resp_wd} as \cite{flekkoy_1995, flekkoy_1996}
\beq \chi_{\kv,\text{wd}}(t) = \frac{k}{2\rho \omega_c} \sin(\omega_c t)\theta(t)\,, \eeq
from which the equal-time height-correlation function can be obtained as
\beq \begin{split}\bra |h_{\kv,\text{wd}}(t)|^2\ket &= \int_0^t ds \int_0^t ds' \chi_{\kv,\text{wd}}(t-s) \chi_{\kv,\text{wd}}(t-s') \bra F_\kv(s)F_\kv^*(s')\ket\\
&= \frac{2 k_B T}{\sigma k^2}\sin^2( \omega_c t)\,.
\end{split}\label{cap_hcorr_wd}\eeq
Due to the neglect of viscosity, the correlation function describes infinitely oscillating capillary waves.
It is useful to note that, at early times ($t\ll 1/w_c$), $\bra |h_\kv(t)|^2\ket$ grows $\propto t^2$ (Fig.~\ref{fig:cap_rough_th}a).
The time-dependent interfacial roughness, eq.~\eqref{cap_rough_def}, follows in 2D from eq.~\eqref{cap_hcorr_wd} as \cite{flekkoy_1996}
\beq W^2\st{wd}(t) = \frac{k_B T}{\pi\sigma} \int^{k\st{max}}_{k\st{min}} dk \frac{\sin^2 ( \omega_c t)}{k^2}
= t^{2/3} \frac{2 k_B T}{3\pi (2\rho \sigma^2)^{1/3} }  \int^{ \omega\st{c,max}t}_{ \omega\st{c,min}t} dx \frac{\sin^2 x}{x^{5/3}}\,,
\label{cap_rough_wd}
\eeq
where $k\st{min}$ and $k\st{max}$ are the smallest and largest possible wavenumbers in the system and $\omega_{c,\text{min,max}}$ are the corresponding capillary wave frequencies. In the last step, the substitution $x= \omega_{c} t$ has been made to exhibit the leading time dependence.
In the range where $\omega\st{c,max}^{-1} \lesssim t\lesssim \omega\st{c,min}^{-1}$, the value of the integral is roughly constant and, consequently, $W\st{wd}(t)$ grows like $t^{1/3}$. At early times, $W\st{wd}(t)$ grows linearly in time (Fig.~\ref{fig:cap_rough_th}b), reflecting the quadratic growth of the height-correlation function [eq.~\eqref{cap_hcorr_wd}]. Note that the extent of the early-time regime decreases relative to the late-time regime when the size $L$ of the interface is increased.
In 3D, we have logarithmic growth, $W^2\st{wd}(t)\propto \log t$ \cite{flekkoy_1996}.
The time after which the roughness reaches its equilibrium value can be estimated as a quarter of a period of the capillary wave with the largest wavelength,
\beq t_{r,\text{wd}} \simeq  \sqrt{L^3\rho/(16\pi\sigma)}\,. \label{cap_tr_wd}
\eeq

In the \textit{strong-damping limit}, the Fourier transform of eq.~\eqref{cap_resp_od} yields
\beq \chi_{\kv,\text{sd}}(t) = \frac{1}{4 \eta k}\exp\left(-\frac{\sigma k}{4\eta}t\right)\theta(t)\,.\eeq
From the expression for the correlations of the random force in this limit,
\beq \bra F_\kv(t) F_\kv^*(0)\ket = 4\rho\nu k k_B T \delta(|t|)\,,
\label{rand_force_sd}
\eeq
the equal-time height-correlation function results as
\beq \bra |h_{\kv,\text{sd}}(t)|^2\ket = \frac{k_B T}{\sigma k^2}\left[1-\exp\left(-\frac{\sigma k}{2\eta}t\right)\right]\,.\label{cap_hcorr_od}\eeq
A factor of 2 has been included in the prefactor of \eqref{cap_hcorr_od} to recover the correct expression for the static spectrum, eq.~\eqref{cap_hcorr_eq}, in the long-time limit.
According to eq.~\eqref{cap_hcorr_od}, the height variance grows $\propto t$ at small times until, after roughly a timescale of the order of the capillary time ($\sim \eta/\sigma k$), it reaches equilibrium (Fig.~\ref{fig:cap_rough_th}a).
In 2D, the roughness follows from eq.~\eqref{cap_hcorr_od} as
\beq W^2\st{sd}(t) = t \frac{k_B T}{4\pi \eta} \int_{a k\st{min}t}^{a k\st{max}t} dx \frac{1-\exp(-x)}{x^2} = t \frac{k_B T}{4\pi \eta} \left[\Gamma(-1,x)-x^{-1}\right]\Big|_{x= a k\st{min}t}^{x= a k\st{max}t}\,,
\label{cap_rough_od}
\eeq
where $a\equiv \sigma/2\eta$ and $\Gamma(n,x)$ is the incomplete Gamma function.
As long as $(a k\st{max})^{-1}\lesssim  t\lesssim (a k\st{min} )^{-1}$, the leading time-dependence is unaffected by the expression in the square brackets and the roughness thus grows as $W\st{sd}(t)\propto t^{1/2}$ until equilibrium is reached. At early times, $W_{\text{sd}}(t)$ is expected to grow linearly, which is related to higher-order frequency terms that are neglected in eq.~\eqref{cap_resp_od} (see appendix for further discussion).
In 3D, we find
\beq W^2\st{sd}(t) = \frac{k_B T}{4\pi\sigma}[-\text{Ei}(-x)+\log x]\Big|_{x= a k\st{min}t}^{x= a k\st{max}t}\,,
\eeq
where $\text{Ei}$ denotes the exponential integral function. In the range $(a k\st{max})^{-1}\lesssim  t\lesssim (a k\st{min} )^{-1}$, $W^2\st{sd}(t)\propto \log t$, which is preceded by a linear growth at earlier times.
The roughening time in the strong-damping regime can be approximated as the inverse of the overdamped relaxation rate of eq.~\eqref{cap_overdmp} evaluated for the largest wavenumber,
\beq t_{r,\text{sd}}\simeq \eta L/\pi\sigma\,.\label{cap_tr_od}
\eeq
Note that, at late times, the normalization of the above $W\st{sd}^2(t)$ is slightly different from the exact equilibrium result based on the sum over the discrete wavemodes, eq.~\eqref{cap_rough_eq}.

The origin of the different scaling behavior for the roughness in eqs.~\eqref{cap_rough_wd} and \eqref{cap_rough_od} lies in the different growth speeds of the individual height amplitudes $\bra |h_\kv(t)|^2\ket$ in dependence of the wavenumber (Fig.~\ref{fig:cap_rough_th}a). In the strong-damping case, the growth rate scales $\propto 1/k$ at early times, i.e., the mode with the smallest wavenumber grows fastest and the roughness essentially reflects the growth of this mode. In contrast, in the weak-damping case, modes of larger $k$ grow faster and the early linear growth of $W$ is essentially due to the largest $k$-mode available in the system. However, since large-$k$ modes also reach equilibrium earlier [from eq.~\eqref{cap_hcorr_wd}, their roughening time is $\propto k^{-3/2}$], there appears a $t^{1/3}$-regime in the weak-damping case where the growth of $W$ is due to a ``saturation'' effect of subsequently smaller $k$-modes, until, finally, also the mode with the smallest $k$ has reached equilibrium.

Neglecting the very early growth, it is seen that, in 2D, the roughness $W(t)$ obeys a scaling relation
\beq  W(t) = L^\alpha w(t/L^z)\,, \text{with } \alpha=1/2,\, \begin{cases}z=3/2 \text{\, (weak damping)}\\z=1 \text{\, (strong damping).}\end{cases} \label{cap_rough_scal_form}\eeq
The scaling index $\alpha$ is an equilibrium property of the roughness [eq.~\eqref{cap_rough_eq}] and is independent of the damping regime.
The scaling function $w(x)$ reaches a constant for $x\ra \infty$ and behaves as 
\beq w(x)\sim x^{\alpha/z}
\eeq 
for small $x$.
As noted in \cite{flekkoy_1996, starr_boghosian_prl1996}, in the weak-damping limit, the scaling properties of $W$ fall into the Kardar-Parisi-Zhang universality class of surface growth \cite{kpz_prl1986}.\footnote{As noted previously \cite{flekkoy_1996}, this might be by coincidence, though, since the growth mechanism described by the Kardar-Parisi-Zhang equation is different from the roughening of a fluid interface.}
However, as shown here, the Langevin theory predicts a different dynamical scaling index $z$ for strong damping.\footnote{Interestingly, in the strong-damping case, the dynamic exponents given in eq.~\eqref{cap_rough_scal_form} can also be derived from the renormalization group study of \cite{bray_long_pre2001}, although the results obtained there are formally valid only for $d>2$.}
In three dimensions, the roughening proceeds logarithmically both in the weak and strong damping regime and can not be characterized by scaling exponents.

\section{Simulations}
\label{sec:sim}
\subsection{Model and setup}
\label{sec:model}
Our simulations are based on the fluctuating Lattice Boltzmann (LB) method \cite{ladd_1994, adhikari_fluct_2005, duenweg_statmechLB_2007}, which was extended to non-ideal fluids in \cite{gross_flbe_2010} (see also \cite{gross_fdbe_2011, gross_critstat_2012} for further discussions).
The underlying deterministic LB model, on which our fluctuating LB approach is based, was introduced in \cite{swift_prl1995, swift_pre1996}.
For general information on the LB method we refer to the numerous reviews available \cite{benzi_physrep1992,raabe_overview_2004,succi_book}. 
For the present purposes, it suffices to note that this method solves the equations of fluctuating hydrodynamics of an isothermal non-ideal fluid \cite{Landau_FluidMech59, sengers_book, hohenberg_halperin, felderhoff_physica1970, langer_turski_pra1973, anderson_diffuse_1998}. 
The fluctuating hydrodynamic equations describe the evolution of a fluctuating density and velocity field $\rho(\rv)$, $\uv(\rv)$ and consist of a continuity equation,
\beq \pd_t \rho = -\nabla\cdot (\rho \uv) \label{cont_eq}\eeq
and a momentum conservation equation,
\beq
\pd_t (\rho \uv) + \nabla\cdot (\rho \uv\uv) = -\nabla\cdot \ten{P} + \nabla\cdot \ten{\Pi} + \nabla\cdot \ten{R}\,. \label{nse}
\eeq
Here, $\ten \Pi$ is the viscous stress tensor 
\beq \Pi_{\alpha\beta} = \eta \left(\pd_\alpha u_\beta + \pd_\beta u_\alpha - \frac{2}{d} \delta_{\alpha\beta} \pd_\gamma u_\gamma\right) + \zeta \pd_\gamma u_\gamma \delta_{\alpha\beta}\,,
\label{visc_ten}
\eeq
$\ten R$ the random stress tensor with Gaussian correlations characterized by
\beq \bra R_{\alpha \beta}(\bv{r},t) R_{\gamma \delta}(\bv{r'},t') \ket = 2k_B T \left[ \eta \left(\delta_{\alpha \gamma} \delta_{\beta \delta} + \delta_{\alpha \delta} \delta_{\beta \gamma} - \frac{2}{d} \delta_{\alpha \beta}\delta_{\gamma \delta}\right)
+ \zeta\, \delta_{\alpha \beta}\delta_{\gamma \delta}\right]\delta(\bv{r}-\bv{r'})\delta(t-t')
\,,
\label{rand_stress}\eeq
and $\eta$ and $\zeta$ the shear and bulk viscosity. 
As a peculiarity of the LB method, $\eta$ and $\zeta$ are in general proportional to the density, but can otherwise be freely tuned.
For generality, we have kept the spatial dimensionality $d$ here.
The physics of a non-ideal fluid enters the model via the pressure tensor
\cite{felderhoff_physica1970, evans_interface_1979, anderson_diffuse_1998}
\beq P_{\alpha\beta} = \left(p_0 - \kappa \rho \nabla^2 \rho - \frac{\kappa}{2}|\nabla\rho|^2\right) \delta_{\alpha\beta} + \kappa (\pd_\alpha \rho) (\pd_\beta \rho)\,,
\label{press_ten}
\eeq
where $p_0$ is a given equation of state (bulk pressure) and $\kappa$ a square-gradient parameter.
In fact, the above form of the pressure tensor $\ten P$ can be derived from a Ginzburg-Landau free energy functional 
\beq
\Fcal[\rho] = \int d\rv\left[ \frac{\kappa}{2}|\nabla \rho|^2 + f_0(\rho) \right]\,,
\label{fef_rho}
\eeq
since
\beq \nabla \cdot \ten{P} = \rho \nabla \frac{\delta \Fcal}{\delta\rho}\,.
\label{pten_force}
\eeq
Here, $f_0$ is a Landau free energy density, fulfilling 
\beq p_0 = \rho \pd_\rho f_0 - f_0\,.
\label{p0}
\eeq
We take $f_0$ to be of a simple double-well form with minima at the equilibrium densities $\rho_V$, $\rho_L$ \cite{RowlinsonWidom_book, Chaikin_book, jamet_second_2001}
\beq  f_0(\rho) = \beta (\rho - \rho_V)^2(\rho-\rho_L)^2\,.
\label{f0}
\eeq
Note that for a symmetric Ginzburg-Landau free energy functional [as in eq.~\eqref{fef_rho}] the bending rigidity -- and with it the Tolman correction, which describes the dependence of the surface tension on curvature \cite{blokhuis_bedeaux_jcp1992} -- vanishes \cite{fisher_wortis_prb1984}.
In the absence of flow and thermal noise, the equilibrium solution of eqs.~\eqref{cont_eq}, \eqref{nse} and \eqref{f0} fulfils $\nabla\cdot \ten P=0$ and describes, in the simplest case, a liquid and a vapor phase separated by a diffuse interface. The profile is given by eq.~\eqref{intprof}, with a width of
\beq w=\frac{2}{\rho_L-\rho_V}\sqrt{\frac{2\kappa}{\beta}}\,.
\eeq
Thermal noise, which is imparted by the random stress tensor $\ten R$ throughout the fluid, leads to the excitation of capillary waves on the interface \cite{felderhoff_physica1970, grant_desai_1983, flekkoy_1995, flekkoy_1996}.
Only a small number of previous simulation studies of fluctuating interfaces in continuum hydrodynamic models exist, of which the ones most relevant for the present context are based on lattice gas automata \cite{flekkoy_1995, flekkoy_1996, starr_boghosian_prl1996} (see also \cite{shang_jcp2011}). Recently, also a fluctuating LB scheme for the Kardar-Parisi-Zhang equation has been introduced \cite{yermakou_kpz_2012}.

All our simulations are performed in two dimensions. We place a liquid stripe of size $L_X\times H$ in a rectangular box of size $L_X\times L_Y$, where $L_Y$ is typically $\sim 2H\simeq 128$ lattice units (l.u.) and $L_X$ varies between 128 and 512 l.u. 
The box is periodic in $x$-direction and covered by substrates at the boundaries in $y$-direction (see Fig.~\ref{fig:capill-def}).
The ratio of the equilibrium roughness to the film thickness is less than $O(10^{-2})$. We expect our simulations to approximate the limit of infinitely deep films and have checked in a few cases, by increasing $H$ and $L_Y$, that our results are insensitive to the film height.
Static properties of capillary fluctuations turn out to be quite insensitive to the specific simulation parameters in the present model. In contrast, regarding dynamics, satisfactory agreement between theory and simulation results was found to require quite large liquid-vapor density ratios (around $\rho_L/\rho_V=100$) and intrinsic interface widths not larger than 5 l.u.

\subsection{Equilibrium properties}

\begin{figure}[t]\centering
    \includegraphics[width=0.43\linewidth]{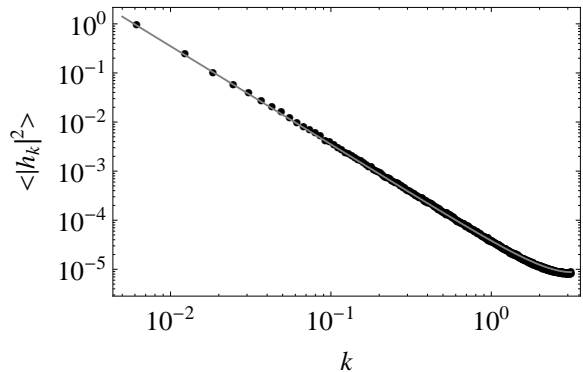}
   \caption{Equal-time spectrum of interfacial height fluctuations on a planar one-dimensional interface obtained from LB simulations (filled circles), compared to the theoretical capillary structure factor [solid line, eq.~\eqref{cap_hcorr_eq}]. $k$ denotes the wavenumber in the plane of the interface. Simulation parameters: $L=1024$, $\rho_L=1.0$, $\rho_V=0.5$, $\beta=0.11$, $\kappa=0.08$, $k_B T = 10^{-7}$, $\tau=1.0$, surface tension $\sigma\simeq 2.7\times 10^{-3}$, interface width approximately 5 l.u.}
    \label{fig:capill-static}
\end{figure}

Before we turn to interfacial roughening, a number of basic equilibrium properties of capillary waves are discussed. This not only serves as a validation of the simulation method, but is also important since, owing to the fluctuation-dissipation relation, equilibrium dynamics and non-equilibrium roughening are governed by the same kinetic coefficients.
In Fig.~\ref{fig:capill-static}, the static (equal-time) capillary wave correlation function $\bra h_\kv h_{-\kv}\ket$ obtained from our simulations is shown (see also \cite{gross_fdbe_2011}). 
The real-space height profile $h(\rv)$ is extracted by fitting the mean-field profile, eq.~\eqref{intprof}, at each point $\rv$ to the instantaneous density profile obtained from the simulation.
The spectrum is computed -- after neglecting the initial roughening period -- by averaging over 2000 snapshots in a simulation running for $10^6$ timesteps, which is around one order of magnitude larger than the largest possible relaxation time of a capillary fluctuation in the system, as inferred from eqs.~\eqref{cap_tr_wd} or \eqref{cap_tr_od} (both of which give similar estimates).
Perfect agreement between the simulation results and the theoretical capillary structure factor, eq.~\eqref{cap_hcorr_eq}, is found for practically all wavenumbers.
Note that, on a lattice, the $k^2$ in eq.~\eqref{cap_hcorr_eq} has to be replaced by (the negative of) the Fourier-transform of the proper one-dimensional discrete Laplacian, $2-2\cos k$. The discrete nature of the Laplacian is the reason for the upturn of the structure factor at large wavenumbers in Fig.~\ref{fig:capill-static}.
The difference between the continuum and lattice Laplacian is significant only for large wavenumbers ($k\gtrsim 1$).

\begin{figure}[t]\centering
    (a)\includegraphics[width=0.43\linewidth]{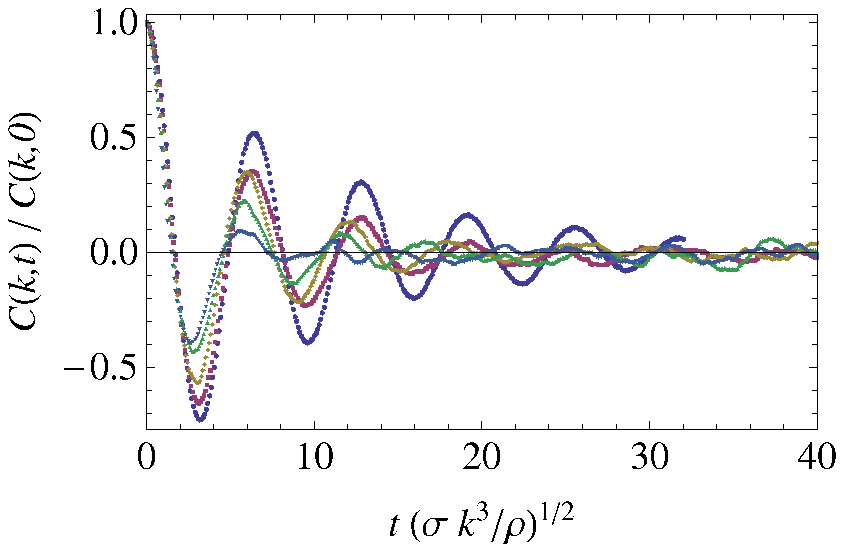}\quad
    (b)\includegraphics[width=0.43\linewidth]{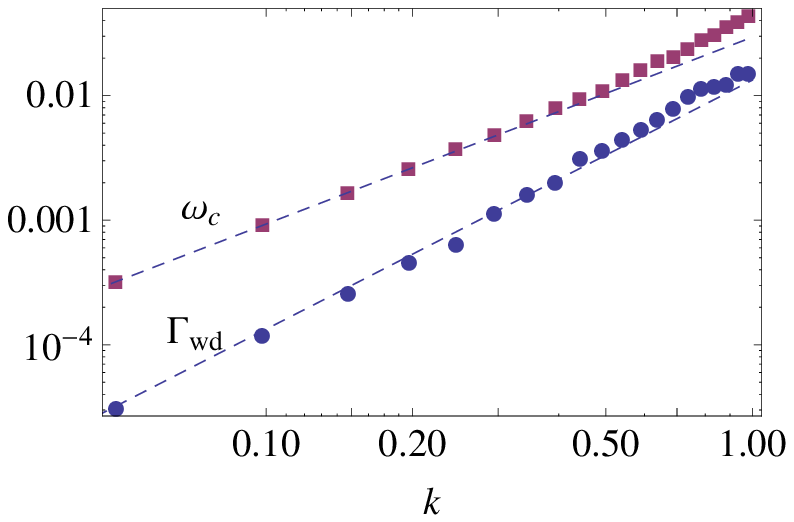}
   \caption{(Color online) Capillary waves on a planar interface in the weak-damping regime. In (a), the correlation function $C(k,t)$ obtained from simulations (normalized to its equilibrium value) is shown for different wavenumbers $k=2\pi n/L$, where $n=1\,(\bullet)$, 3 ({\tiny$\blacksquare$}), 5 ({\small$\blacklozenge$}), 7 ($\blacktriangle$), 9 ($\blacktriangledown$). The time axis is scaled by the theoretical resonance frequency $w_c$, eq.~\eqref{cap_disp}. In (b), the data for the capillary wave resonance frequency $\omega_c(k)$ ({\tiny$\blacksquare$}) and damping rate $\Gamma(k)$ ($\bullet$) are shown. The dashed lines represent the theoretical predictions of eq.~\eqref{cap_disp}, $\sqrt{2}\omega_c$ and $2\Gamma\st{wd}$, corrected by numerical prefactors. Simulation parameters: $L=128$, $\rho_L=1.0$, $\rho_V=0.01$, $\beta=0.0024$, $\kappa=0.006$, $\nu=0.00667$, $k_B T = 10^{-8}$, surface tension $\sigma=8.7\times 10^{-4}$, interface width $\simeq$ 5 l.u.}
    \label{fig:capill_wd}
\end{figure}

\begin{figure}[t]\centering
    (a)\includegraphics[width=0.44\linewidth]{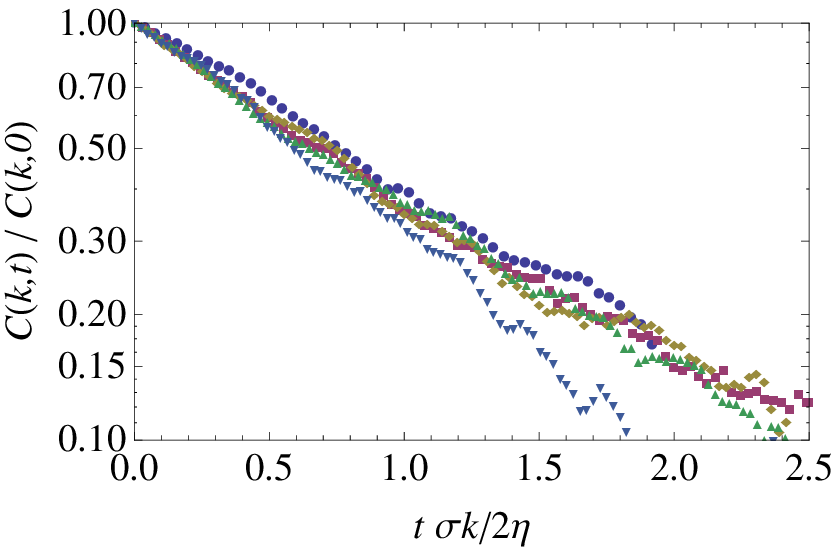}\quad
    (b)\includegraphics[width=0.45\linewidth]{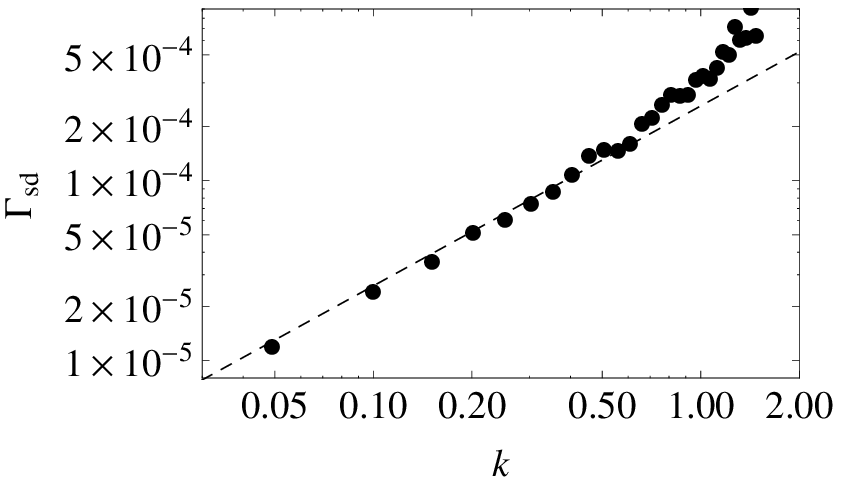}
   \caption{(Color online) Capillary waves on a planar interface in the strong-damping regime. In (a), the correlation function $C(k,t)$ obtained from simulations (normalized to its equilibrium value) is shown for different wavenumbers $k=2\pi n/L$, where $n=1\,(\bullet)$, 3 ({\tiny$\blacksquare$}), 5 ({\small$\blacklozenge$}), 7 ($\blacktriangle$), 9 ($\blacktriangledown$). The time axis is scaled by the theoretical relaxation rate, eq.~\eqref{cap_overdmp}. In (b), the data for the damping rate in the overdamped regime are shown. The dashed line represents the theoretical prediction of eq.~\eqref{cap_overdmp}, $2\Gamma_{sd}$, corrected by a numerical prefactor. Simulation parameters: $L=128$, $\rho_L=1.0$, $\rho_V=0.01$, $\beta=0.00024$, $\kappa=0.0006$, $\nu=0.167$, $k_B T = 10^{-9}$, surface tension $\sigma=8.7\times 10^{-5}$, interface width $\simeq$ 5 l.u.}
    \label{fig:capill_od}
\end{figure}

Fig.~\ref{fig:capill_wd}a shows simulation results on the capillary wave correlation function $C(\kv,t) \equiv \bra h_\kv(t) h_{-\kv}\ket$ in the weak-damping regime (see caption to Fig.~\ref{fig:capill_wd} for simulation parameters), where an oscillatory decay,
\beq C(\kv,t) = \bra | h_\kv |^2\ket \exp(-\Gamma\st{wd} |t|) \cos(\omega_c t)\,, \label{cap_wd_decay}\eeq
with a frequency and damping rate given by eq.~\eqref{cap_disp}, is predicted by the theory.
Since the damping rate increases quadratically with $k$, waves with larger wavenumbers are seen to oscillate for not more than a single period until they become practically indistinguishable from the background noise. 
Fig.~\ref{fig:capill_wd}b shows the oscillation frequency and damping rate extracted by fitting expression \eqref{cap_wd_decay} to the simulation data. The dashed lines represent $\sqrt{2}\omega_c$ and $2\Gamma\st{wd}$, where the prefactors have been included here in order to conform with the expressions of standard capillary wave theories. It is seen that, after the correction for numerical prefactors, both quantities compare well to the theoretical predictions up to a wavenumber of $k\simeq 0.7$. Deviations at higher wavenumbers are partly attributed to the fact that the damping is so strong that it is hard to unambiguously extract both the frequency and damping rate from the data. Additionally, the hydrodynamic regime, where the transport coefficients are constants, generally breaks down in LB at large wavenumbers \cite{behrend_hydro_1994, lallemand_theory_2000}.

In the case of overdamped dynamics, capillary waves decay purely exponentially:
\beq C(\kv,t) = \bra |h_\kv|^2\ket \exp(-\Gamma\st{sd} |t|) \,,\label{cap_od_decay}\eeq
which is clearly seen in the logarithmic plot in Fig.~\ref{fig:capill_od}a. 
Fig.~\ref{fig:capill_od}b shows the relaxation rate, obtained by fitting eq.~\eqref{cap_od_decay} to the data. After correcting for a numerical prefactor of 2, the theoretical prediction is well reproduced up to a wavenumber of $k\simeq 0.5$.
Similarly to the previous case, a possible reason for the noticeable deviations at larger wavenumbers might be that the viscosity becomes wavenumber-dependent for larger $k$.
It should also be noted that the above results are obtained in the limit of small vapor density ($\rho_V=0.01\rho_L$). For larger vapor densities, the agreement between simulation and theoretical predictions is found to become worse, even when comparing to theoretical expressions that take into account a finite vapor density \cite{bouchiat_meunier_jphys1974, jeng_capvisc_1998}. We also observe that the discrepancies grow when the interfacial width is further increased.
The origin of this behavior is presently unknown.
Thus, in the future, a closer theoretical investigation of capillary wave dynamics in diffuse interface models would be interesting, taking also into account effects of a finite vapor density, which seems to have not been done in a sufficiently general way up to now \cite{felderhoff_physica1970, turski_langer_pra1980}.

Of course, since the above results pertain to the linear-response regime (where the fluctuation amplitudes are small by definition), they could have equivalently been obtained from a study of individual capillary waves in a simulation without thermal noise (Onsager regression hypothesis).
Previous LB studies of capillary wave dynamics made use of this equivalence and obtained capillary wave dispersion relations similar to the present work \cite{shan_chen_1994, swift_prl1995,  zhang_cpp2000}.
In contrast to the equilibrium relaxation dynamics, however, the interfacial roughening phenomenon studied in the next section is a genuinely fluctuation induced effect.

\subsection{Interfacial roughening}
\begin{figure}[t]\centering
    (a)\includegraphics[width=0.43\linewidth]{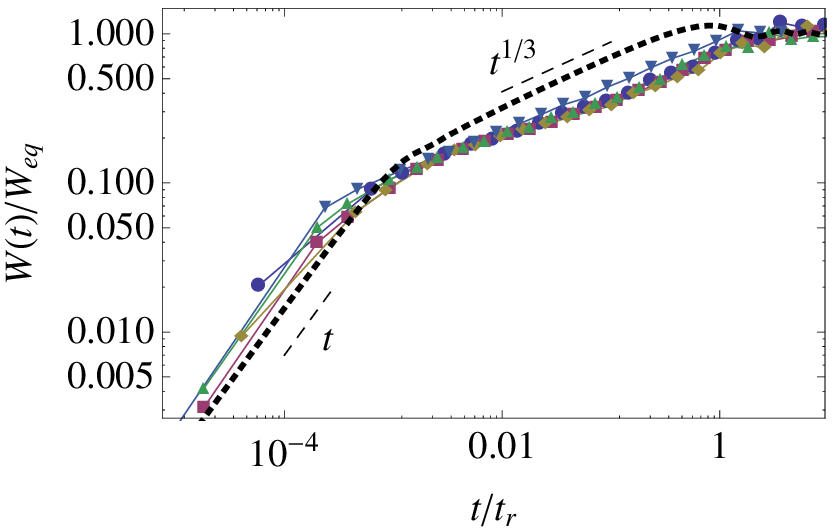}
    (b)\includegraphics[width=0.43\linewidth]{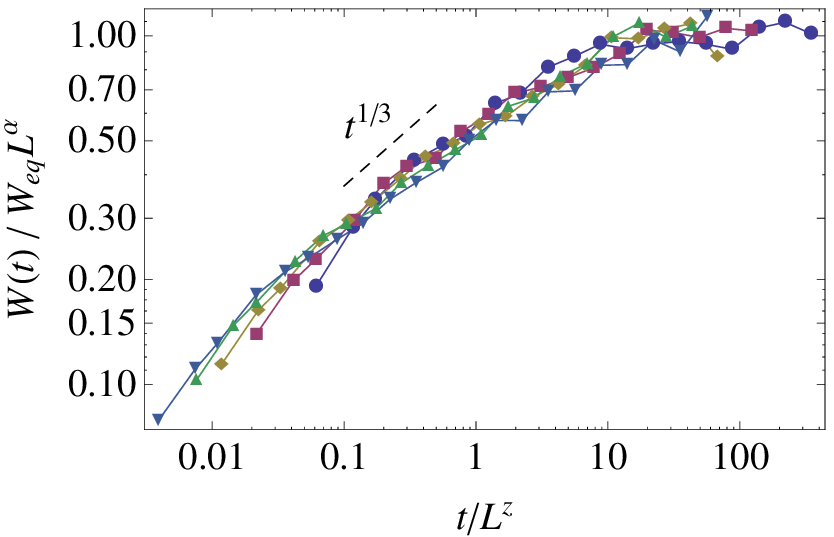}
    \caption{(Color online) Time-evolution of the interfacial roughness $W$ in the \textit{weak-damping} regime. In (a), data obtained for different values of surface tension, viscosity and fluctuation temperature (see Table~\ref{tab:sim_rough}) and a fixed system size of $L=512$ are plotted. The thick dotted curve represents the theoretical prediction, eq.~\eqref{cap_rough_wd}. Time is scaled by the roughening time $t_r$ [eq.~\eqref{cap_tr_wd}] and the roughness is scaled by its equilibrium value $W\st{eq}$. In (b), simulation parameters are fixed at $\rho_L=1.0$, $\rho_V=0.01$, $\sigma=8.7\times 10^{-4}$, $\nu=6.7\times 10^{-3}$, $k_B T=10^{-8}$ and the system size is varied as $L=32\,(\bullet)$, 64 ({\tiny$\blacksquare$}), 96 ({\small$\blacklozenge$}), 128 ($\blacktriangle$), 200 ($\blacktriangledown$). The values of the scaling exponents are $\alpha=1/2$ and $z=3/2$.}
    \label{fig:cap_rough_wd}
\end{figure}

\begin{figure}[t]\centering
    (a)\includegraphics[width=0.43\linewidth]{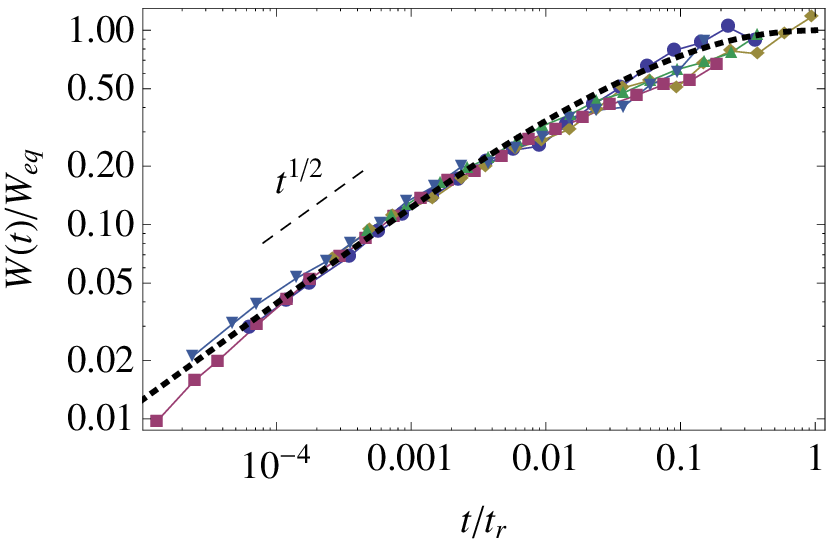}
    (b)\includegraphics[width=0.43\linewidth]{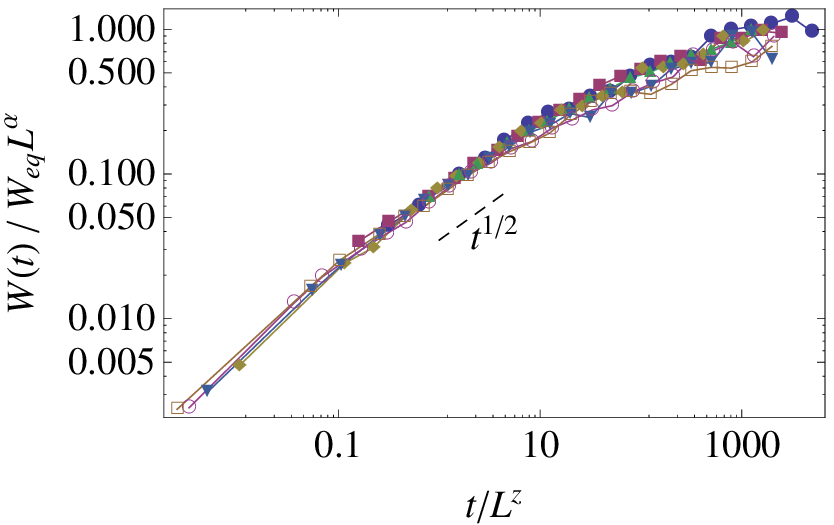}
    \caption{(Color online) Time-evolution of the interfacial roughness $W$ in the \textit{strong-damping} regime. In (a), data obtained for different values of surface tension, viscosity and fluctuation temperature (see Table~\ref{tab:sim_rough}) and a fixed system size of $L=128$ are plotted. The thick dotted line represents the theoretical prediction, eq.~\eqref{cap_rough_od}. Time is scaled by the roughening time $t_r$ [eq.~\eqref{cap_tr_od}] and the roughness is scaled by its equilibrium value $W\st{eq}$. In (b), simulation parameters are fixed at $\rho_L=1.0$, $\rho_V=0.01$, $\sigma=3.6\times 10^{-5}$, $\nu=0.17$, $k_B T=10^{-9}$ and the system size is varied as $L=32\,(\bullet)$, 64 ({\tiny$\blacksquare$}), 96 ({\small$\blacklozenge$}), 128 ($\blacktriangle$), 200 ($\blacktriangledown$), 300 ($\circ$), 400 ({\tiny$\square$}). The values of the scaling exponents are $\alpha=1/2$ and $z=1$.}
    \label{fig:cap_rough_od}
\end{figure}

\begin{table}[b]
\begin{center}
\begin{tabular}{c | c c c | c c c}
\hline\hline
    & \multicolumn{3}{c|}{weak-damping (Fig.~\ref{fig:cap_rough_wd}a)} & \multicolumn{3}{c}{strong-damping (Fig.~\ref{fig:cap_rough_od}a)} \\
    & $\sigma$ & $\nu$ & $k_B T$ &  $\sigma$ & $\nu$ & $k_B T$ \\
\hline
$\bullet$ & $8.7\times 10^{-3}$ & $0.017$ & $10^{-8}$ & $8.7\times 10^{-5}$ & 0.17 & $10^{-9}$ \\
{\tiny$\blacksquare$} & $8.7\times 10^{-4}$ & $3.3\times 10^{-3}$ & $10^{-9}$ & $3.6\times 10^{-6}$ & 0.033 & $10^{-10}$ \\
{\small$\blacklozenge$} & $4.3\times 10^{-3}$ & $6.7\times 10^{-3}$ & $10^{-8}$ & $7.2\times 10^{-4}$ & 0.33 & $10^{-10}$ \\
$\blacktriangle$ & $8.7\times 10^{-4}$ & $6.7\times 10^{-3}$ & $10^{-8}$ & $3.6\times 10^{-5}$ & 0.17 & $10^{-9}$ \\
$\blacktriangledown$ & $1.4\times 10^{-5}$ & $5.0 \times 10^{-3}$ & $10^{-10}$ & $3.6\times 10^{-6}$ & 0.17 & $10^{-10}$ \\
\hline\hline
\end{tabular}
\end{center}
\caption{Simulation parameters (surface tension $\sigma$, kinematic shear viscosity $\nu$ and fluctuation temperature $k_B T$) used in Figs.~\ref{fig:cap_rough_wd}a and \ref{fig:cap_rough_od}a. In both cases, $\rho_L=1.0$ and $\rho_V=0.01$. All parameters are given in l.u..
}
\label{tab:sim_rough}
\end{table}

Figures~\ref{fig:cap_rough_wd} and \ref{fig:cap_rough_od} show simulation results for the time-evolution of the interfacial roughness in the weak- and strong-damping regimes. It has been made sure, by choosing simulation parameters appropriately (see Table~\ref{tab:sim_rough}), that all wavemodes existing on the interface exclusively fall in either one of the considered regimes. In all cases, the interfacial width is 5 l.u.\ and the density ratio of $\rho_L/\rho_V=100$ is used in order to approximate the case of zero vapor density (upon which the theoretical derivation is based) as closely as possible. To ensure sufficient statistical accuracy, the roughness is computed by averaging over $10-30$ independent simulations.
Since it was seen in Fig.~\ref{fig:capill-static} that the static capillary correlations are correctly reproduced, it is clear that the equilibrium interfacial roughness [eq.~\eqref{cap_rough_eq}] -- which is essentially determined by the integrated structure factor -- also agrees with the theoretical predictions. Therefore, $W\eq$ is not discussed separately here, but instead, we turn directly to the time-evolution of the roughness.

In Fig.~\ref{fig:cap_rough_wd}a, the time-dependent interfacial roughness $W(t)$ in the weak-damping case is plotted, with each curve corresponding to a simulation performed for different values of surface tension, viscosity and fluctuation temperature, keeping the system size fixed. Data collapse is achieved by rescaling time by the roughening time $t_{r,\text{wd}}$, eq.~\eqref{cap_tr_wd}, and the roughness by its expected equilibrium value $W\eq$, eq.~\eqref{cap_rough_eq}.
We see that the overall trend of the data is correctly captured by the theory (dashed curve), with a linear growth at early times and a $t^{1/3}$-growth at late times, until at around a time $t_{r,\text{wd}}$ the roughness attains its equilibrium value.
In the crossover region between the two growth regimes, however, the roughness is found to grow significantly slower than predicted by the theory. In fact, the data seem to be more consistent with a $t^{1/4}$ behavior at intermediate times. This effect is more pronounced for small system sizes and also found to slightly depend on the chosen simulation parameters.

\begin{figure}[t]\centering
    \includegraphics[width=0.48\linewidth]{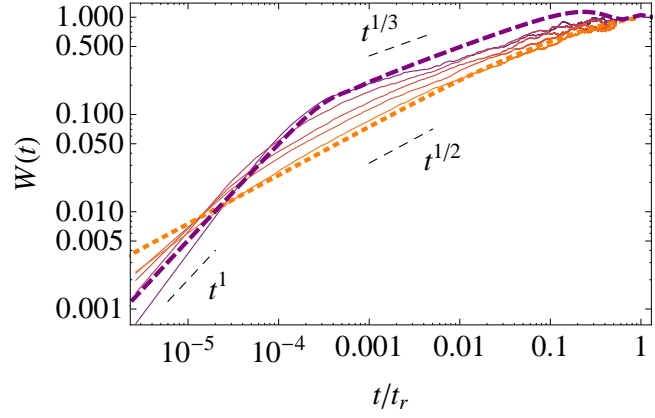}
    \caption{(Color online) Time-evolution of the roughness in the crossover region between weak and strong damping. The dashed and dotted curves represent the expressions for $W$ in the weak and strong-damping limit, eq.~\eqref{cap_rough_wd} and eq.~\eqref{cap_rough_od}, respectively. Simulation data are shown as thin curves for visibility, corresponding to viscosities of $\nu=0.0033, 0.0167, 0.04, 0.067, 0.133$ l.u. The other simulation parameters are fixed at $\rho_L=1.0$, $\rho_V=0.01$, $\sigma=1.4\times 10^{-5}$, $k_B T=10^{-10}$, $L=128$. Time is rescaled by the roughening time corresponding to the strong-damping limit, eq.~\eqref{cap_tr_od}.}
    \label{fig:rough_crossover}
\end{figure}

In Fig.~\ref{fig:cap_rough_wd}b, the roughness is shown for different system sizes between $L=32$ and $200$ l.u., keeping all other system parameters the same. By scaling time with $L^z$ ($z=3/2$) and $W(t)$ with $L^\alpha$ ($\alpha=1/2$), it is seen that all data points approximately collapse onto a single master curve with a logarithmic slope of $\alpha/z=1/3$, as expected from the scaling form \eqref{cap_rough_scal_form}.
However, we remark that a satisfactory scaling collapse could also be achieved with an index of $z=2$ (or any other value between $3/2$ and 2), corresponding to a growth behavior $W\sim t^{1/4}$ in Fig.~\ref{fig:cap_rough_wd}a. 
While the reason for these discrepancies between simulation and theory is unclear at present, we note that these deviations cannot be explained by the influence of a finite viscosity, since in that case one would expect to approach a $t^{1/2}$-power-law with increasing viscosity and thus, find an even larger exponent than $1/3$ (cf.~Fig.~\ref{fig:rough_crossover}). Also, the effect seems not to be related to the density ratio between liquid and vapor, as we have obtained essentially the same results for different values of $\rho_L/\rho_V$ ranging between 0.5 and~0.001.

Fig.~\ref{fig:cap_rough_od} shows the time-evolution of the interfacial roughness in the strong-damping regime. In Fig.~\ref{fig:cap_rough_od}a, the data are obtained from simulations of varying fluid parameters but identical system size, while in \ref{fig:cap_rough_od}b only the system size is varied.
Good agreement with the theoretical predictions [eq.~\eqref{cap_rough_od}] over approximately three orders of magnitude is found. In particular, the data agree well with the derived scaling function (dashed curve in \ref{fig:cap_rough_od}a) and the approximate power-law growth of the roughness $\propto t^{1/2}$ is recovered. At early times, some deviations from a pure power-law behavior are visible, which can be attributed to the neglect of terms beyond linear order in the frequency-dependence of the response function [see eq.~\eqref{cap_resp_od}]. This point is further discussed in appendix \ref{app:early_time}. 
In contrast to the weak-damping case, no rescaling of time in the plot of the theoretical $W(t)$ is found to be necessary. The scaling collapse of the data in Fig.~\ref{fig:cap_rough_od}b is achieved with a dynamical index $z=1$, confirming that the roughening dynamics in the overdamped case belongs to a different universality class than in the weak-damping case.

Finally, we investigate in Fig.~\ref{fig:rough_crossover} the roughness in the crossover region from weak to strong damping. The simulation data (thin solid lines) in Fig.~\ref{fig:rough_crossover} have been obtained by successively increasing the viscosity from small to large values, keeping all other system parameters fixed. Due to the non-Markovian nature of the interface dynamics in the crossover regime, it is difficult to obtain an analytic expression for $W(t)$ in this case. We observe that the simulation data smoothly interpolate between the overdamped (thick dotted line) and underdamped (thick dashed line) limits. Similarly to Fig.~\ref{fig:cap_rough_wd}a, deviations between simulations and theory are noticeable at intermediate times in the limit of weak-damping (they appear to be more pronounced here due to the smaller system size than in Fig.~\ref{fig:cap_rough_wd}a).

\section{Summary}
We have investigated in this work the dynamics of capillary waves and the non-equilibrium roughening of a liquid-vapor interface based on an effective Langevin description and by means of fluctuating hydrodynamics simulations using the Lattice Boltzmann method. 
Although roughening is a well-known mechanism of film growth, its counterpart in thermally excited fluid interfaces seems to have been only rarely studied \cite{flekkoy_1995, flekkoy_1996, starr_boghosian_prl1996, bray_long_pre2001}. 
As a central result, we showed that the non-equilibrium growth of the roughness proceeds by different dynamical scaling laws, depending on whether the system is either weakly or strongly damped [see eq.~\eqref{cap_rough_scal_form}]. 
In the weak-damping case, we find basic agreement between our Lattice-Boltzmann simulations and previous works \cite{flekkoy_1995, flekkoy_1996,starr_boghosian_prl1996}, which observed scaling exponents characteristic for the Kardar-Parisi-Zhang universality class ($\alpha=1/2$, $z=3/2$). We remark, however, that an dynamic exponent of $z=2$ can also not be fully excluded based on the present data. 
In the strong damping case, the roughening is governed by a dynamic scaling exponent $z=1$, which is characteristic for an overdamped harmonic oscillator driven by white noise.
The scaling exponent $\alpha$, related to the size dependence of the equilibrium roughness, is found to be equal to $1/2$, in agreement with the theory.

\begin{acknowledgements}
We would like to thank A. J. Wagner for useful discussions. Funding from the industrial sponsors of ICAMS, the state of North-Rhine Westphalia and the European Commission in the framework of the European Regional Development Fund (ERDF) is gratefully acknowledged.
\end{acknowledgements}


\appendix
\section{Early-time behavior}
\label{app:early_time}
\begin{figure}[t]\centering
    (a)\includegraphics[width=0.43\linewidth]{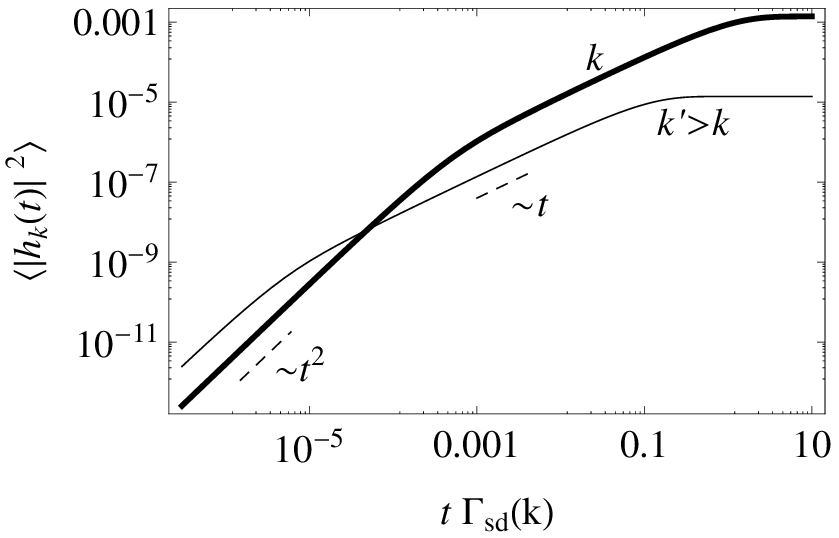}
    (b)\includegraphics[width=0.43\linewidth]{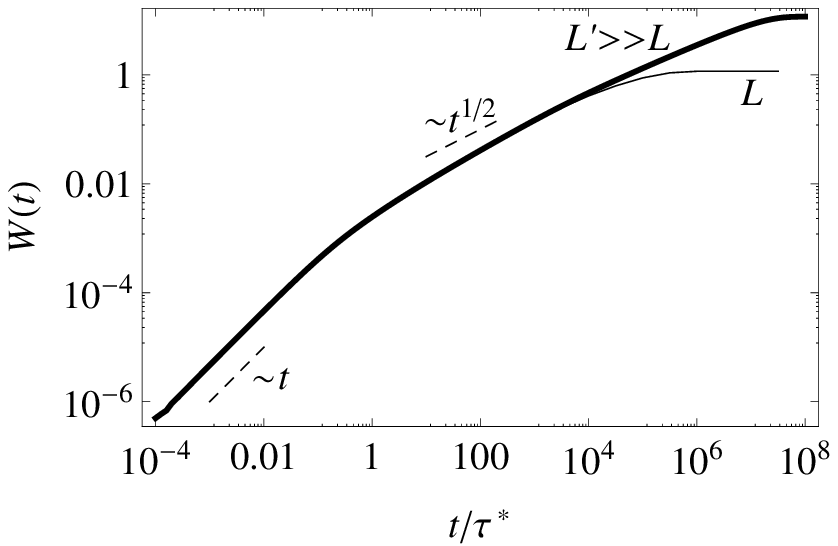}
    \caption{Early-time dynamics in the overdamped regime. (a) Equal-time height-correlation function, taking into account the leading-order frequency dependence [eq.~\eqref{hcorr_sd_ext}], for two different wavenumbers $k$ and $k'>k$. Time is scaled by the damping rate [eq.~\eqref{cap_overdmp}] corresponding to the wavenumber $k$. (b) Time-evolution of the roughness based on expression \eqref{hcorr_sd_ext}, for two different system sizes $L$ and $L'=100L$. Time is scaled by the crossover time $\tau^*=1/k\st{max}^2\nu$ between early- and late-time behavior.}
    \label{fig:cap_rough_ext}
\end{figure}

Relation \eqref{cap_resp_od} and thus eq.~\eqref{cap_rough_od} for the roughness in the case of strong damping becomes exact in the limit of infinite viscosity. For large but finite viscosity, eq.~\eqref{cap_rough_od} describes the growth only at sufficiently late times, since, by keeping only the leading term in the expansion of \eqref{cap_gamma_od}, the high-frequency aspects of the dynamics have been neglected. In the case of weak damping, we expect the early-time properties to be correctly described by eq.~\eqref{cap_rough_wd}, since the leading term in the expansion of eq.~\eqref{cap_gamma_wd} is already the dominant one for large frequencies. Some insights into the early-time growth of a strongly damped interface can be gained by retaining the leading frequency-dependent term in $\gamma\st{sd}$. This results in a harmonic-oscillator form of the response function
\beq 
\chi_{\kv,\text{sd}}(\omega)= \frac{k}{3\rho}\; \frac{1}{-\omega^2-\frac{4}{3}\nu k^2 \im\omega + \sigma k^3/3\rho}
\label{response_od_ext}
\eeq
which becomes in real space
\beq
\chi_{\kv,\text{sd}}(t)=  \frac{k}{3\rho D} \exp\left(-\frac{2}{3}\nu k^2 t\right) 2 \sinh\left(\onehalf D t\right)\,,
\eeq
with $D\equiv \sqrt{16 \nu^2 k^4/9 - 4\sigma k^3/3\rho}$.
The correlations of the random force acquire now, in addition to expression \eqref{rand_force_sd}, a contribution proportional to the derivative of a $\delta$-function,
\beq \bra F_\kv(t) F^*_\kv(0)\ket =k_B T \left[4\rho\nu k \delta(t) + \frac{3\rho}{k} \delta'(t)\right]\,.
\eeq
When computing the height correlation function, this term is treated as in the weak-damping case \cite{flekkoy_1996}. 
In this way, we find
\beq \bra |h_\kv(t)|^2\ket = \frac{k_B T}{2\sigma k^2}\left[1- \exp\left(-\frac{4}{3}k^2\nu t\right)\left( \cosh D t - \frac{4 \nu  k^2}{3D} \sinh D t \right) \right]\,.
\label{hcorr_sd_ext}
\eeq
The correlation function \eqref{hcorr_sd_ext} is plotted in Fig.~\ref{fig:cap_rough_ext}a for two different wavenumbers.
The early-time growth of the height-correlation function can be more directly assessed  by neglecting the ``mass term'' $\sigma k^3/3\rho$ in the response function \eqref{response_od_ext}, yielding
\beq
\bra |h_\kv(t)|^2\ket \simeq \frac{k_B T}{32 k^3 \nu^2 \rho}\left[8k^2 \nu t + 3 \exp\left(-\frac{8}{3} k^2\nu t\right)-3\right ]\,.
\eeq
Thus, for small times, the height correlation function grows 
$\propto k t^2 $, until for $t\gtrsim (\nu k^2)^{-1}$, crossover to linear growth, characteristic of the overdamped case in the infinite-viscosity limit (Brownian dynamics), occurs. 
As an artifact of neglecting the surface tension, the system roughens for an infinite time.
Note that since the frequency-expansion of $\gamma\st{sd}$ has been truncated after the linear order term [eq.~\eqref{cap_gamma_od}], a different growth behavior might result at still earlier times. 

Figure~\ref{fig:cap_rough_ext}b shows the time-dependent roughness obtained by integrating eq.~\eqref{hcorr_sd_ext} over all wavenumbers of a finite system. The early-time growth of $\bra |h_\kv(t)|^2\ket$ is directly reflected in the linear growth of $W(t)$, while the $t^{1/2}$-growth characteristic for an overdamped system sets in after a crossover time $\tau^*$. We find that $\tau^*\simeq (k\st{max}^2\nu)^{-1}$ is determined only by the viscosity and the largest wavenumber of the system (here, $k\st{max}=\pi/l_0$), but is independent of the system size. Thus, for $\nu\ra \infty$ or $L\ra \infty$, the growth of the roughness will be completely dominated by the late-time behavior, where the scaling expressed by eq.~\eqref{cap_rough_scal_form} holds.


\begin{thebibliography}{69}
\expandafter\ifx\csname natexlab\endcsname\relax\def\natexlab#1{#1}\fi
\expandafter\ifx\csname bibnamefont\endcsname\relax
  \def\bibnamefont#1{#1}\fi
\expandafter\ifx\csname bibfnamefont\endcsname\relax
  \def\bibfnamefont#1{#1}\fi
\expandafter\ifx\csname citenamefont\endcsname\relax
  \def\citenamefont#1{#1}\fi
\expandafter\ifx\csname url\endcsname\relax
  \def\url#1{\texttt{#1}}\fi
\expandafter\ifx\csname urlprefix\endcsname\relax\def\urlprefix{URL }\fi
\providecommand{\bibinfo}[2]{#2}
\providecommand{\eprint}[2][]{\url{#2}}

\bibitem[{\citenamefont{Mandelstam}(1913)}]{mandelstam_1913}
\bibinfo{author}{\bibfnamefont{L.}~\bibnamefont{Mandelstam}},
  \bibinfo{journal}{Ann. Phys. (Leipzig)} \textbf{\bibinfo{volume}{41}},
  \bibinfo{pages}{608} (\bibinfo{year}{1913}).

\bibitem[{\citenamefont{Buff et~al.}(1965)\citenamefont{Buff, Lovett, and
  {Stillinger Jr.}}}]{buff_1965}
\bibinfo{author}{\bibfnamefont{F.~P.} \bibnamefont{Buff}},
  \bibinfo{author}{\bibfnamefont{R.~A.} \bibnamefont{Lovett}},
  \bibnamefont{and} \bibinfo{author}{\bibfnamefont{F.~H.}
  \bibnamefont{{Stillinger Jr.}}}, \bibinfo{journal}{Phys. Rev. Lett.}
  \textbf{\bibinfo{volume}{15}}, \bibinfo{pages}{621} (\bibinfo{year}{1965}).

\bibitem[{\citenamefont{Evans}(1979)}]{evans_interface_1979}
\bibinfo{author}{\bibfnamefont{R.}~\bibnamefont{Evans}}, \bibinfo{journal}{Adv.
  Phys.} \textbf{\bibinfo{volume}{28}}, \bibinfo{pages}{143}
  (\bibinfo{year}{1979}).

\bibitem[{\citenamefont{Rowlinson and Widom}(1982)}]{RowlinsonWidom_book}
\bibinfo{author}{\bibfnamefont{J.~S.} \bibnamefont{Rowlinson}}
  \bibnamefont{and} \bibinfo{author}{\bibfnamefont{B.}~\bibnamefont{Widom}},
  \emph{\bibinfo{title}{{Molecular Theory of Capillarity}}}
  (\bibinfo{publisher}{Dover Publications}, \bibinfo{year}{1982}).

\bibitem[{\citenamefont{Safran}(1994)}]{safran_book}
\bibinfo{author}{\bibfnamefont{S.~A.} \bibnamefont{Safran}},
  \emph{\bibinfo{title}{Statistical Thermodynamics of Surfaces, Interfaces and
  Membranes}} (\bibinfo{publisher}{Addison-Wesley Publishing},
  \bibinfo{year}{1994}).

\bibitem[{\citenamefont{Loudon}(1980)}]{loudon_1980}
\bibinfo{author}{\bibfnamefont{R.}~\bibnamefont{Loudon}},
  \bibinfo{journal}{Proc. R. Soc. Lond. A} \textbf{\bibinfo{volume}{372}},
  \bibinfo{pages}{275} (\bibinfo{year}{1980}).

\bibitem[{\citenamefont{Grant and Desai}(1983)}]{grant_desai_1983}
\bibinfo{author}{\bibfnamefont{M.}~\bibnamefont{Grant}} \bibnamefont{and}
  \bibinfo{author}{\bibfnamefont{R.~C.} \bibnamefont{Desai}},
  \bibinfo{journal}{Phys. Rev. A} \textbf{\bibinfo{volume}{27}},
  \bibinfo{pages}{2577} (\bibinfo{year}{1983}).

\bibitem[{\citenamefont{Family and Vicsek}(1991)}]{family_viscek_book1991}
\bibinfo{author}{\bibfnamefont{F.}~\bibnamefont{Family}} \bibnamefont{and}
  \bibinfo{author}{\bibfnamefont{T.}~\bibnamefont{Vicsek}},
  \emph{\bibinfo{title}{Dynamics of fractal surfaces}}
  (\bibinfo{publisher}{World Scientific}, \bibinfo{year}{1991}).

\bibitem[{\citenamefont{Barabasi and
  Stanley}(1995)}]{barabasi_stanley_fractal_book}
\bibinfo{author}{\bibfnamefont{A.-L.} \bibnamefont{Barabasi}} \bibnamefont{and}
  \bibinfo{author}{\bibfnamefont{H.~E.} \bibnamefont{Stanley}},
  \emph{\bibinfo{title}{{Fractal Concepts in Surface Growth}}}
  (\bibinfo{publisher}{Cambridge University Press, Cambridge},
  \bibinfo{year}{1995}).

\bibitem[{\citenamefont{Krug and Spohn}(1991)}]{krug_spohn_roughening_1991}
\bibinfo{author}{\bibfnamefont{J.}~\bibnamefont{Krug}} \bibnamefont{and}
  \bibinfo{author}{\bibfnamefont{H.}~\bibnamefont{Spohn}}, in
  \emph{\bibinfo{booktitle}{Solids far from equilibrium}}, edited by
  \bibinfo{editor}{\bibfnamefont{C.}~\bibnamefont{Godreche}}
  (\bibinfo{publisher}{Cambridge University Press, Cambridge},
  \bibinfo{year}{1991}), p. \bibinfo{pages}{479}.

\bibitem[{\citenamefont{Meakin}(1993)}]{meakin_physrep1993}
\bibinfo{author}{\bibfnamefont{P.}~\bibnamefont{Meakin}},
  \bibinfo{journal}{Phys. Rep.} \textbf{\bibinfo{volume}{235}},
  \bibinfo{pages}{189} (\bibinfo{year}{1993}).

\bibitem[{\citenamefont{Halpin-Healy and
  Zhang}(1995)}]{halpinhealy_zhang_roughening_1995}
\bibinfo{author}{\bibfnamefont{T.}~\bibnamefont{Halpin-Healy}}
  \bibnamefont{and} \bibinfo{author}{\bibfnamefont{Y.}~\bibnamefont{Zhang}},
  \bibinfo{journal}{Phys. Rep.} \textbf{\bibinfo{volume}{254}},
  \bibinfo{pages}{215} (\bibinfo{year}{1995}).

\bibitem[{\citenamefont{Edwards and Wilkinson}(1982)}]{edwards_wilkinson_1982}
\bibinfo{author}{\bibfnamefont{S.~F.} \bibnamefont{Edwards}} \bibnamefont{and}
  \bibinfo{author}{\bibfnamefont{D.~R.} \bibnamefont{Wilkinson}},
  \bibinfo{journal}{Proc. R. Soc. London} \textbf{\bibinfo{volume}{381}},
  \bibinfo{pages}{17} (\bibinfo{year}{1982}).

\bibitem[{\citenamefont{Kardar et~al.}(1986)\citenamefont{Kardar, Parisi, and
  Zhang}}]{kpz_prl1986}
\bibinfo{author}{\bibfnamefont{M.}~\bibnamefont{Kardar}},
  \bibinfo{author}{\bibfnamefont{G.}~\bibnamefont{Parisi}}, \bibnamefont{and}
  \bibinfo{author}{\bibfnamefont{Y.-C.} \bibnamefont{Zhang}},
  \bibinfo{journal}{Phys. Rev. Lett.} \textbf{\bibinfo{volume}{56}},
  \bibinfo{pages}{889} (\bibinfo{year}{1986}).

\bibitem[{\citenamefont{Flekkoy and Rothman}(1995)}]{flekkoy_1995}
\bibinfo{author}{\bibfnamefont{E.~G.} \bibnamefont{Flekkoy}} \bibnamefont{and}
  \bibinfo{author}{\bibfnamefont{D.~H.} \bibnamefont{Rothman}},
  \bibinfo{journal}{Phys. Rev. Lett.} \textbf{\bibinfo{volume}{75}},
  \bibinfo{pages}{260} (\bibinfo{year}{1995}).

\bibitem[{\citenamefont{Flekkoy and Rothman}(1996)}]{flekkoy_1996}
\bibinfo{author}{\bibfnamefont{E.~G.} \bibnamefont{Flekkoy}} \bibnamefont{and}
  \bibinfo{author}{\bibfnamefont{D.~H.} \bibnamefont{Rothman}},
  \bibinfo{journal}{Phys. Rev. E} \textbf{\bibinfo{volume}{53}},
  \bibinfo{pages}{1622} (\bibinfo{year}{1996}).

\bibitem[{\citenamefont{Foard and
  Wagner}(2012)}]{foard_wagner_morphologies_pre2012}
\bibinfo{author}{\bibfnamefont{E.~M.} \bibnamefont{Foard}} \bibnamefont{and}
  \bibinfo{author}{\bibfnamefont{A.~J.} \bibnamefont{Wagner}},
  \bibinfo{journal}{Phys. Rev. E} \textbf{\bibinfo{volume}{85}},
  \bibinfo{pages}{011501} (\bibinfo{year}{2012}).

\bibitem[{\citenamefont{Ayodele et~al.}(2011)\citenamefont{Ayodele, Varnik, and
  Raabe}}]{ayodele_grayscott_pre2011}
\bibinfo{author}{\bibfnamefont{S.}~\bibnamefont{Ayodele}},
  \bibinfo{author}{\bibfnamefont{F.}~\bibnamefont{Varnik}}, \bibnamefont{and}
  \bibinfo{author}{\bibfnamefont{D.}~\bibnamefont{Raabe}},
  \bibinfo{journal}{Phys. Rev. E} \textbf{\bibinfo{volume}{83}},
  \bibinfo{pages}{016702} (\bibinfo{year}{2011}).

\bibitem[{\citenamefont{Wagner and
  Yeomans}(1999)}]{wagner_yeomans_shear_pre1999}
\bibinfo{author}{\bibfnamefont{A.~J.} \bibnamefont{Wagner}} \bibnamefont{and}
  \bibinfo{author}{\bibfnamefont{J.~M.} \bibnamefont{Yeomans}},
  \bibinfo{journal}{Phys. Rev. E} \textbf{\bibinfo{volume}{59}},
  \bibinfo{pages}{4366} (\bibinfo{year}{1999}).

\bibitem[{\citenamefont{Shou and Chakrabarti}(2000)}]{shou_chakrabarti_pre2000}
\bibinfo{author}{\bibfnamefont{Z.}~\bibnamefont{Shou}} \bibnamefont{and}
  \bibinfo{author}{\bibfnamefont{A.}~\bibnamefont{Chakrabarti}},
  \bibinfo{journal}{Phys. Rev. E} \textbf{\bibinfo{volume}{61}},
  \bibinfo{pages}{R2200} (\bibinfo{year}{2000}).

\bibitem[{\citenamefont{Bray et~al.}(2001{\natexlab{a}})\citenamefont{Bray,
  Cavagna, and Travasso}}]{bray_short_pre2001}
\bibinfo{author}{\bibfnamefont{A.~J.} \bibnamefont{Bray}},
  \bibinfo{author}{\bibfnamefont{A.}~\bibnamefont{Cavagna}}, \bibnamefont{and}
  \bibinfo{author}{\bibfnamefont{R.~D.~M.} \bibnamefont{Travasso}},
  \bibinfo{journal}{Phys. Rev. E} \textbf{\bibinfo{volume}{64}},
  \bibinfo{pages}{012102} (\bibinfo{year}{2001}{\natexlab{a}}).

\bibitem[{\citenamefont{Bray et~al.}(2001{\natexlab{b}})\citenamefont{Bray,
  Cavagna, and Travasso}}]{bray_long_pre2001}
\bibinfo{author}{\bibfnamefont{A.~J.} \bibnamefont{Bray}},
  \bibinfo{author}{\bibfnamefont{A.}~\bibnamefont{Cavagna}}, \bibnamefont{and}
  \bibinfo{author}{\bibfnamefont{R.~D.~M.} \bibnamefont{Travasso}},
  \bibinfo{journal}{Phys. Rev. E} \textbf{\bibinfo{volume}{65}},
  \bibinfo{pages}{016104} (\bibinfo{year}{2001}{\natexlab{b}}).

\bibitem[{\citenamefont{Aarts and
  Lekkerkerker}(2008)}]{aarts_coalescence_jfm2008}
\bibinfo{author}{\bibfnamefont{D.~G. A.~L.} \bibnamefont{Aarts}}
  \bibnamefont{and} \bibinfo{author}{\bibfnamefont{H.~N.~W.}
  \bibnamefont{Lekkerkerker}}, \bibinfo{journal}{J. Fluid. Mech.}
  \textbf{\bibinfo{volume}{606}}, \bibinfo{pages}{275} (\bibinfo{year}{2008}).

\bibitem[{\citenamefont{Grant}(1988)}]{grant_wetting_prb1988}
\bibinfo{author}{\bibfnamefont{M.}~\bibnamefont{Grant}},
  \bibinfo{journal}{Phys. Rev. B} \textbf{\bibinfo{volume}{37}},
  \bibinfo{pages}{5705} (\bibinfo{year}{1988}).

\bibitem[{\citenamefont{Starr et~al.}(1996)\citenamefont{Starr, Harrington,
  Boghosian, and Stanley}}]{starr_boghosian_prl1996}
\bibinfo{author}{\bibfnamefont{F.~W.} \bibnamefont{Starr}},
  \bibinfo{author}{\bibfnamefont{S.~T.} \bibnamefont{Harrington}},
  \bibinfo{author}{\bibfnamefont{B.~M.} \bibnamefont{Boghosian}},
  \bibnamefont{and} \bibinfo{author}{\bibfnamefont{H.~E.}
  \bibnamefont{Stanley}}, \bibinfo{journal}{Phys. Rev. Lett.}
  \textbf{\bibinfo{volume}{77}}, \bibinfo{pages}{3363} (\bibinfo{year}{1996}).

\bibitem[{\citenamefont{Zittartz}(1967)}]{zittartz_1967}
\bibinfo{author}{\bibfnamefont{J.}~\bibnamefont{Zittartz}},
  \bibinfo{journal}{Phys. Rev.} \textbf{\bibinfo{volume}{154}},
  \bibinfo{pages}{154} (\bibinfo{year}{1967}).

\bibitem[{\citenamefont{Diehl et~al.}(1980)\citenamefont{Diehl, Kroll, and
  Wagner}}]{diehl_zphysb1980}
\bibinfo{author}{\bibfnamefont{H.~W.} \bibnamefont{Diehl}},
  \bibinfo{author}{\bibfnamefont{D.~M.} \bibnamefont{Kroll}}, \bibnamefont{and}
  \bibinfo{author}{\bibfnamefont{H.}~\bibnamefont{Wagner}},
  \bibinfo{journal}{Z. Phys. B} \textbf{\bibinfo{volume}{36}},
  \bibinfo{pages}{329} (\bibinfo{year}{1980}).

\bibitem[{\citenamefont{Evans}(1981)}]{evans_1981}
\bibinfo{author}{\bibfnamefont{R.}~\bibnamefont{Evans}}, \bibinfo{journal}{Mol.
  Phys.} \textbf{\bibinfo{volume}{42}}, \bibinfo{pages}{1169}
  (\bibinfo{year}{1981}).

\bibitem[{\citenamefont{Stecki}(1998)}]{stecki_contrib_1998}
\bibinfo{author}{\bibfnamefont{J.}~\bibnamefont{Stecki}}, \bibinfo{journal}{J.
  Chem. Phys.} \textbf{\bibinfo{volume}{108}}, \bibinfo{pages}{3788}
  (\bibinfo{year}{1998}).

\bibitem[{\citenamefont{Sedlmeier et~al.}(2009)\citenamefont{Sedlmeier,
  Horinek, and Netz}}]{sedlmeier_netz_prl2009}
\bibinfo{author}{\bibfnamefont{F.}~\bibnamefont{Sedlmeier}},
  \bibinfo{author}{\bibfnamefont{D.}~\bibnamefont{Horinek}}, \bibnamefont{and}
  \bibinfo{author}{\bibfnamefont{R.~R.} \bibnamefont{Netz}},
  \bibinfo{journal}{Phys. Rev. Lett.} \textbf{\bibinfo{volume}{103}},
  \bibinfo{pages}{136102} (\bibinfo{year}{2009}).

\bibitem[{\citenamefont{Blokhuis}(2009)}]{blokhuis_jcp2009}
\bibinfo{author}{\bibfnamefont{E.~M.} \bibnamefont{Blokhuis}},
  \bibinfo{journal}{J. Chem. Phys.} \textbf{\bibinfo{volume}{130}},
  \bibinfo{pages}{014706} (\bibinfo{year}{2009}).

\bibitem[{\citenamefont{Jeng et~al.}(1998)\citenamefont{Jeng, Esibov, Crow, and
  Steyerl}}]{jeng_capvisc_1998}
\bibinfo{author}{\bibfnamefont{U.-S.} \bibnamefont{Jeng}},
  \bibinfo{author}{\bibfnamefont{L.}~\bibnamefont{Esibov}},
  \bibinfo{author}{\bibfnamefont{L.}~\bibnamefont{Crow}}, \bibnamefont{and}
  \bibinfo{author}{\bibfnamefont{A.}~\bibnamefont{Steyerl}},
  \bibinfo{journal}{J. Phys.: Cond. Mat.} \textbf{\bibinfo{volume}{10}},
  \bibinfo{pages}{4955} (\bibinfo{year}{1998}).

\bibitem[{\citenamefont{Madsen et~al.}(2004)\citenamefont{Madsen, Seydel,
  Sprung, Gutt, Tolan, and Gr\"{u}bel}}]{madsen_prl2004}
\bibinfo{author}{\bibfnamefont{A.}~\bibnamefont{Madsen}},
  \bibinfo{author}{\bibfnamefont{T.}~\bibnamefont{Seydel}},
  \bibinfo{author}{\bibfnamefont{M.}~\bibnamefont{Sprung}},
  \bibinfo{author}{\bibfnamefont{C.}~\bibnamefont{Gutt}},
  \bibinfo{author}{\bibfnamefont{M.}~\bibnamefont{Tolan}}, \bibnamefont{and}
  \bibinfo{author}{\bibfnamefont{G.}~\bibnamefont{Gr\"{u}bel}},
  \bibinfo{journal}{Phys. Rev. Lett.} \textbf{\bibinfo{volume}{92}},
  \bibinfo{pages}{096104} (\bibinfo{year}{2004}).

\bibitem[{\citenamefont{Harden et~al.}(1991)\citenamefont{Harden, Pleiner, and
  Pincus}}]{harden_pleiner_pincus_jcp1991}
\bibinfo{author}{\bibfnamefont{J.~L.} \bibnamefont{Harden}},
  \bibinfo{author}{\bibfnamefont{H.}~\bibnamefont{Pleiner}}, \bibnamefont{and}
  \bibinfo{author}{\bibfnamefont{P.~A.} \bibnamefont{Pincus}},
  \bibinfo{journal}{J. Chem. Phys.} \textbf{\bibinfo{volume}{94}},
  \bibinfo{pages}{5208} (\bibinfo{year}{1991}).

\bibitem[{\citenamefont{Jaeckle and
  Kawasaki}(1995)}]{jaeckle_kawasaki_jpcm1995}
\bibinfo{author}{\bibfnamefont{J.}~\bibnamefont{Jaeckle}} \bibnamefont{and}
  \bibinfo{author}{\bibfnamefont{K.}~\bibnamefont{Kawasaki}},
  \bibinfo{journal}{J. Phys.: Cond. Mat.} \textbf{\bibinfo{volume}{7}},
  \bibinfo{pages}{4351} (\bibinfo{year}{1995}).

\bibitem[{\citenamefont{Thiebaud and Bickel}(2010)}]{thiebaud_shear_pre2010}
\bibinfo{author}{\bibfnamefont{M.}~\bibnamefont{Thiebaud}} \bibnamefont{and}
  \bibinfo{author}{\bibfnamefont{T.}~\bibnamefont{Bickel}},
  \bibinfo{journal}{Phys. Rev. E} \textbf{\bibinfo{volume}{81}},
  \bibinfo{pages}{031602} (\bibinfo{year}{2010}).

\bibitem[{\citenamefont{Levich}(1962)}]{levich_book_1962}
\bibinfo{author}{\bibfnamefont{V.}~\bibnamefont{Levich}},
  \emph{\bibinfo{title}{Physicochemical hydrodynamics}}
  (\bibinfo{publisher}{Prentice Hall}, \bibinfo{year}{1962}).

\bibitem[{\citenamefont{Bouchiat and
  Meunier}(1972)}]{bouchiat_meunier_jphys1972}
\bibinfo{author}{\bibfnamefont{M.~A.} \bibnamefont{Bouchiat}} \bibnamefont{and}
  \bibinfo{author}{\bibfnamefont{J.}~\bibnamefont{Meunier}},
  \bibinfo{journal}{J. de Phys.} \textbf{\bibinfo{volume}{33}},
  \bibinfo{pages}{C1} (\bibinfo{year}{1972}).

\bibitem[{\citenamefont{Mora and Daillant}(2002)}]{mora_daillant_epjb2002}
\bibinfo{author}{\bibfnamefont{S.}~\bibnamefont{Mora}} \bibnamefont{and}
  \bibinfo{author}{\bibfnamefont{J.}~\bibnamefont{Daillant}},
  \bibinfo{journal}{Eur. J. Phys. B} \textbf{\bibinfo{volume}{27}},
  \bibinfo{pages}{417} (\bibinfo{year}{2002}).

\bibitem[{\citenamefont{Felderhoff}(1970)}]{felderhoff_physica1970}
\bibinfo{author}{\bibfnamefont{B.~U.} \bibnamefont{Felderhoff}},
  \bibinfo{journal}{Physica A} \textbf{\bibinfo{volume}{48}},
  \bibinfo{pages}{541} (\bibinfo{year}{1970}).

\bibitem[{\citenamefont{Bouchiat and
  Meunier}(1974)}]{bouchiat_meunier_jphys1974}
\bibinfo{author}{\bibfnamefont{M.~A.} \bibnamefont{Bouchiat}} \bibnamefont{and}
  \bibinfo{author}{\bibfnamefont{J.}~\bibnamefont{Meunier}},
  \bibinfo{journal}{J. de Phys.} \textbf{\bibinfo{volume}{35}},
  \bibinfo{pages}{847} (\bibinfo{year}{1974}).

\bibitem[{\citenamefont{Falk and Mecke}(2011)}]{falk_mecke_jpcm2011}
\bibinfo{author}{\bibfnamefont{K.}~\bibnamefont{Falk}} \bibnamefont{and}
  \bibinfo{author}{\bibfnamefont{K.}~\bibnamefont{Mecke}}, \bibinfo{journal}{J.
  Phys.: Cond. Mat.} \textbf{\bibinfo{volume}{23}}, \bibinfo{pages}{184103}
  (\bibinfo{year}{2011}).

\bibitem[{\citenamefont{Ladd}(1994)}]{ladd_1994}
\bibinfo{author}{\bibfnamefont{A.~J.~C.} \bibnamefont{Ladd}},
  \bibinfo{journal}{J. Fluid Mech.} \textbf{\bibinfo{volume}{271}},
  \bibinfo{pages}{285} (\bibinfo{year}{1994}).

\bibitem[{\citenamefont{Adhikari et~al.}(2005)\citenamefont{Adhikari,
  Stratford, Cates, and Wagner}}]{adhikari_fluct_2005}
\bibinfo{author}{\bibfnamefont{R.}~\bibnamefont{Adhikari}},
  \bibinfo{author}{\bibfnamefont{K.}~\bibnamefont{Stratford}},
  \bibinfo{author}{\bibfnamefont{M.~E.} \bibnamefont{Cates}}, \bibnamefont{and}
  \bibinfo{author}{\bibfnamefont{A.~J.} \bibnamefont{Wagner}},
  \bibinfo{journal}{Europhys. Lett.} \textbf{\bibinfo{volume}{71}},
  \bibinfo{pages}{473} (\bibinfo{year}{2005}).

\bibitem[{\citenamefont{D\"{u}nweg et~al.}(2007)\citenamefont{D\"{u}nweg,
  Schiller, and Ladd}}]{duenweg_statmechLB_2007}
\bibinfo{author}{\bibfnamefont{B.}~\bibnamefont{D\"{u}nweg}},
  \bibinfo{author}{\bibfnamefont{U.~D.} \bibnamefont{Schiller}},
  \bibnamefont{and} \bibinfo{author}{\bibfnamefont{A.~J.~C.}
  \bibnamefont{Ladd}}, \bibinfo{journal}{Phys. Rev. E}
  \textbf{\bibinfo{volume}{76}}, \bibinfo{pages}{036704}
  (\bibinfo{year}{2007}).

\bibitem[{\citenamefont{Gross et~al.}(2010)\citenamefont{Gross, Adhikari,
  Cates, and Varnik}}]{gross_flbe_2010}
\bibinfo{author}{\bibfnamefont{M.}~\bibnamefont{Gross}},
  \bibinfo{author}{\bibfnamefont{R.}~\bibnamefont{Adhikari}},
  \bibinfo{author}{\bibfnamefont{M.~E.} \bibnamefont{Cates}}, \bibnamefont{and}
  \bibinfo{author}{\bibfnamefont{F.}~\bibnamefont{Varnik}},
  \bibinfo{journal}{Phys. Rev. E} \textbf{\bibinfo{volume}{82}},
  \bibinfo{pages}{056714} (\bibinfo{year}{2010}).

\bibitem[{\citenamefont{Gross et~al.}(2011)\citenamefont{Gross, Cates, Varnik,
  and Adhikari}}]{gross_fdbe_2011}
\bibinfo{author}{\bibfnamefont{M.}~\bibnamefont{Gross}},
  \bibinfo{author}{\bibfnamefont{M.~E.} \bibnamefont{Cates}},
  \bibinfo{author}{\bibfnamefont{F.}~\bibnamefont{Varnik}}, \bibnamefont{and}
  \bibinfo{author}{\bibfnamefont{R.}~\bibnamefont{Adhikari}},
  \bibinfo{journal}{J. Stat. Mech.} \textbf{\bibinfo{volume}{2011}},
  \bibinfo{pages}{P03030} (\bibinfo{year}{2011}).

\bibitem[{\citenamefont{Gross and Varnik}(2012)}]{gross_critstat_2012}
\bibinfo{author}{\bibfnamefont{M.}~\bibnamefont{Gross}} \bibnamefont{and}
  \bibinfo{author}{\bibfnamefont{F.}~\bibnamefont{Varnik}},
  \bibinfo{journal}{Phys. Rev. E} \textbf{\bibinfo{volume}{85}},
  \bibinfo{pages}{056707} (\bibinfo{year}{2012}).

\bibitem[{\citenamefont{Swift et~al.}(1995)\citenamefont{Swift, Osborn, and
  Yeomans}}]{swift_prl1995}
\bibinfo{author}{\bibfnamefont{M.~R.} \bibnamefont{Swift}},
  \bibinfo{author}{\bibfnamefont{W.~R.} \bibnamefont{Osborn}},
  \bibnamefont{and} \bibinfo{author}{\bibfnamefont{J.~M.}
  \bibnamefont{Yeomans}}, \bibinfo{journal}{Phys. Rev. Lett.}
  \textbf{\bibinfo{volume}{75}}, \bibinfo{pages}{830} (\bibinfo{year}{1995}).

\bibitem[{\citenamefont{Swift et~al.}(1996)\citenamefont{Swift, Orlandini,
  Osborn, and Yeomans}}]{swift_pre1996}
\bibinfo{author}{\bibfnamefont{M.~R.} \bibnamefont{Swift}},
  \bibinfo{author}{\bibfnamefont{E.}~\bibnamefont{Orlandini}},
  \bibinfo{author}{\bibfnamefont{W.~R.} \bibnamefont{Osborn}},
  \bibnamefont{and} \bibinfo{author}{\bibfnamefont{J.~M.}
  \bibnamefont{Yeomans}}, \bibinfo{journal}{Phys. Rev. E}
  \textbf{\bibinfo{volume}{54}}, \bibinfo{pages}{5041} (\bibinfo{year}{1996}).

\bibitem[{\citenamefont{Benzi et~al.}(1992)\citenamefont{Benzi, Succi, and
  Vergassola}}]{benzi_physrep1992}
\bibinfo{author}{\bibfnamefont{R.}~\bibnamefont{Benzi}},
  \bibinfo{author}{\bibfnamefont{S.}~\bibnamefont{Succi}}, \bibnamefont{and}
  \bibinfo{author}{\bibfnamefont{M.}~\bibnamefont{Vergassola}},
  \bibinfo{journal}{Phys. Rep.} \textbf{\bibinfo{volume}{222}},
  \bibinfo{pages}{145} (\bibinfo{year}{1992}).

\bibitem[{\citenamefont{Raabe}(2004)}]{raabe_overview_2004}
\bibinfo{author}{\bibfnamefont{D.}~\bibnamefont{Raabe}},
  \bibinfo{journal}{Model. Simul. Mater. Sci. Eng.}
  \textbf{\bibinfo{volume}{12}}, \bibinfo{pages}{R13} (\bibinfo{year}{2004}).

\bibitem[{\citenamefont{Succi}(2001)}]{succi_book}
\bibinfo{author}{\bibfnamefont{S.}~\bibnamefont{Succi}},
  \emph{\bibinfo{title}{The {L}attice {B}oltzmann Equation for Fluid Dynamics
  and Beyond}} (\bibinfo{publisher}{OUP, Oxford}, \bibinfo{year}{2001}).

\bibitem[{\citenamefont{Landau and Lifshitz}(1959)}]{Landau_FluidMech59}
\bibinfo{author}{\bibfnamefont{L.~D.} \bibnamefont{Landau}} \bibnamefont{and}
  \bibinfo{author}{\bibfnamefont{E.~M.} \bibnamefont{Lifshitz}},
  \emph{\bibinfo{title}{Fluid Mechanics}} (\bibinfo{publisher}{Pergamon},
  \bibinfo{year}{1959}).

\bibitem[{\citenamefont{de~Zarate and Sengers}(2006)}]{sengers_book}
\bibinfo{author}{\bibfnamefont{J.~M.~O.} \bibnamefont{de~Zarate}}
  \bibnamefont{and} \bibinfo{author}{\bibfnamefont{J.~V.}
  \bibnamefont{Sengers}}, \emph{\bibinfo{title}{{Hydrodynamic Fluctuations in
  Fluids and Fluid Mixtures}}} (\bibinfo{publisher}{Elsevier},
  \bibinfo{year}{2006}).

\bibitem[{\citenamefont{Hohenberg and Halperin}(1977)}]{hohenberg_halperin}
\bibinfo{author}{\bibfnamefont{P.~C.} \bibnamefont{Hohenberg}}
  \bibnamefont{and} \bibinfo{author}{\bibfnamefont{B.~I.}
  \bibnamefont{Halperin}}, \bibinfo{journal}{Rev. Mod. Phys.}
  \textbf{\bibinfo{volume}{49}}, \bibinfo{pages}{435} (\bibinfo{year}{1977}).

\bibitem[{\citenamefont{Langer and Turski}(1973)}]{langer_turski_pra1973}
\bibinfo{author}{\bibfnamefont{J.~S.} \bibnamefont{Langer}} \bibnamefont{and}
  \bibinfo{author}{\bibfnamefont{L.~A.} \bibnamefont{Turski}},
  \bibinfo{journal}{Phys. Rev. A} \textbf{\bibinfo{volume}{8}},
  \bibinfo{pages}{3230} (\bibinfo{year}{1973}).

\bibitem[{\citenamefont{Anderson et~al.}(1998)\citenamefont{Anderson, McFadden,
  and Wheeler}}]{anderson_diffuse_1998}
\bibinfo{author}{\bibfnamefont{D.~M.} \bibnamefont{Anderson}},
  \bibinfo{author}{\bibfnamefont{G.~B.} \bibnamefont{McFadden}},
  \bibnamefont{and} \bibinfo{author}{\bibfnamefont{A.~A.}
  \bibnamefont{Wheeler}}, \bibinfo{journal}{Annu. Rev. Fluid. Mech.}
  \textbf{\bibinfo{volume}{30}}, \bibinfo{pages}{139} (\bibinfo{year}{1998}).

\bibitem[{\citenamefont{Chaikin and Lubensky}(1995)}]{Chaikin_book}
\bibinfo{author}{\bibfnamefont{P.~M.} \bibnamefont{Chaikin}} \bibnamefont{and}
  \bibinfo{author}{\bibfnamefont{T.~C.} \bibnamefont{Lubensky}},
  \emph{\bibinfo{title}{Principles of Condensed Matter Physics}}
  (\bibinfo{publisher}{Cambridge}, \bibinfo{year}{1995}).

\bibitem[{\citenamefont{Jamet et~al.}(2001)\citenamefont{Jamet, Lebaigue,
  Coutris, and Delhaye}}]{jamet_second_2001}
\bibinfo{author}{\bibfnamefont{D.}~\bibnamefont{Jamet}},
  \bibinfo{author}{\bibfnamefont{O.}~\bibnamefont{Lebaigue}},
  \bibinfo{author}{\bibfnamefont{N.}~\bibnamefont{Coutris}}, \bibnamefont{and}
  \bibinfo{author}{\bibfnamefont{J.~M.} \bibnamefont{Delhaye}},
  \bibinfo{journal}{J. Comp. Phys.} \textbf{\bibinfo{volume}{169}},
  \bibinfo{pages}{624} (\bibinfo{year}{2001}).

\bibitem[{\citenamefont{Blokhuis and Bedeaux}(1992)}]{blokhuis_bedeaux_jcp1992}
\bibinfo{author}{\bibfnamefont{E.~M.} \bibnamefont{Blokhuis}} \bibnamefont{and}
  \bibinfo{author}{\bibfnamefont{D.}~\bibnamefont{Bedeaux}},
  \bibinfo{journal}{J. Chem. Phys.} \textbf{\bibinfo{volume}{97}},
  \bibinfo{pages}{3576} (\bibinfo{year}{1992}).

\bibitem[{\citenamefont{Fisher and Wortis}(1984)}]{fisher_wortis_prb1984}
\bibinfo{author}{\bibfnamefont{M.~P.~A.} \bibnamefont{Fisher}}
  \bibnamefont{and} \bibinfo{author}{\bibfnamefont{M.}~\bibnamefont{Wortis}},
  \bibinfo{journal}{Phys. Rev. B} \textbf{\bibinfo{volume}{29}},
  \bibinfo{pages}{6252} (\bibinfo{year}{1984}).

\bibitem[{\citenamefont{Shang et~al.}(2011)\citenamefont{Shang, Voulgarakis,
  and Chu}}]{shang_jcp2011}
\bibinfo{author}{\bibfnamefont{B.~Z.} \bibnamefont{Shang}},
  \bibinfo{author}{\bibfnamefont{N.~K.} \bibnamefont{Voulgarakis}},
  \bibnamefont{and} \bibinfo{author}{\bibfnamefont{J.-W.} \bibnamefont{Chu}},
  \bibinfo{journal}{J. Chem. Phys.} \textbf{\bibinfo{volume}{135}},
  \bibinfo{pages}{044111} (\bibinfo{year}{2011}).

\bibitem[{\citenamefont{Yermakou and Succi}(2012)}]{yermakou_kpz_2012}
\bibinfo{author}{\bibfnamefont{V.}~\bibnamefont{Yermakou}} \bibnamefont{and}
  \bibinfo{author}{\bibfnamefont{S.}~\bibnamefont{Succi}},
  \bibinfo{journal}{Physica A} \textbf{\bibinfo{volume}{391}}, \bibinfo{pages}{4557} (\bibinfo{year}{2012}).

\bibitem[{\citenamefont{Behrend et~al.}(1994)\citenamefont{Behrend, Harris, and
  Warren}}]{behrend_hydro_1994}
\bibinfo{author}{\bibfnamefont{O.}~\bibnamefont{Behrend}},
  \bibinfo{author}{\bibfnamefont{R.}~\bibnamefont{Harris}}, \bibnamefont{and}
  \bibinfo{author}{\bibfnamefont{P.~B.} \bibnamefont{Warren}},
  \bibinfo{journal}{Phys. Rev. E} \textbf{\bibinfo{volume}{50}},
  \bibinfo{pages}{4586} (\bibinfo{year}{1994}).

\bibitem[{\citenamefont{Lallemand and Luo}(2000)}]{lallemand_theory_2000}
\bibinfo{author}{\bibfnamefont{P.}~\bibnamefont{Lallemand}} \bibnamefont{and}
  \bibinfo{author}{\bibfnamefont{L.-S.} \bibnamefont{Luo}},
  \bibinfo{journal}{Phys. Rev. E} \textbf{\bibinfo{volume}{61}},
  \bibinfo{pages}{6546} (\bibinfo{year}{2000}).

\bibitem[{\citenamefont{Turski and Langer}(1980)}]{turski_langer_pra1980}
\bibinfo{author}{\bibfnamefont{L.~A.} \bibnamefont{Turski}} \bibnamefont{and}
  \bibinfo{author}{\bibfnamefont{J.~S.} \bibnamefont{Langer}},
  \bibinfo{journal}{Phys. Rev. A} \textbf{\bibinfo{volume}{22}},
  \bibinfo{pages}{2189} (\bibinfo{year}{1980}).

\bibitem[{\citenamefont{Shan and Chen}(1994)}]{shan_chen_1994}
\bibinfo{author}{\bibfnamefont{X.}~\bibnamefont{Shan}} \bibnamefont{and}
  \bibinfo{author}{\bibfnamefont{H.}~\bibnamefont{Chen}},
  \bibinfo{journal}{Phys. Rev. E} \textbf{\bibinfo{volume}{49}},
  \bibinfo{pages}{2941} (\bibinfo{year}{1994}).

\bibitem[{\citenamefont{Zhang et~al.}(2000)\citenamefont{Zhang, He, and
  Chen}}]{zhang_cpp2000}
\bibinfo{author}{\bibfnamefont{R.}~\bibnamefont{Zhang}},
  \bibinfo{author}{\bibfnamefont{X.}~\bibnamefont{He}}, \bibnamefont{and}
  \bibinfo{author}{\bibfnamefont{S.}~\bibnamefont{Chen}},
  \bibinfo{journal}{Comp. Phys. Comm.} \textbf{\bibinfo{volume}{129}},
  \bibinfo{pages}{121} (\bibinfo{year}{2000}).

\end{thebibliography}

\end{document}